\documentclass[11pt]{article}
\usepackage{float}
\usepackage{placeins}
\usepackage{amsmath}
\usepackage{amssymb}
\usepackage{amsthm}
\usepackage{amsfonts}
\usepackage{comment}
\usepackage{color}
\usepackage{mathrsfs}
\usepackage{braket}
\usepackage{graphicx}
\usepackage[subrefformat=parens]{subcaption}
\usepackage{cite}
\usepackage{here}
\usepackage{tikz}
\usetikzlibrary{decorations.pathmorphing}
\usepackage{caption}
\usepackage{multirow}
\usepackage[labelfont=bf]{caption}
\usepackage{array} 
\usepackage{hyperref}
\usepackage{listings}
\usepackage{graphicx}
\usepackage{multirow}
\usepackage{array}
\usepackage{geometry}
\usepackage{rotating}
\DeclareFontFamily{U}{stix2scr}{\skewchar\font127 }
\DeclareFontShape{U}{stix2scr}{m}{n} {<-> stix2-mathscr}{}
\DeclareMathAlphabet{\curly}{U}{stix2scr}{m}{n}
\SetMathAlphabet{\curly}{bold}{U}{stix2scr}{b}{n}
\DeclareFontShape{U}{stix2scr}{b}{n} {<-> stix2-mathscr-bold}{}
\DeclareFontShape{U}{stix2scr}{bx}{n} {<->ssub * stix2scr/b/n}{}

\renewcommand{\ell}{\curly{l}}

\usepackage[stable]{footmisc}

\makeatletter

\@addtoreset{equation}{section}
\makeatother

\begin{document}

\pagestyle{empty}

\title{\bf{Grey-body factor from correspondence with quasinormal mode for gravitational perturbation of higher-dimensional Reissner–Nordstr\"{o}m black hole}}

\date{}
\maketitle
\begin{center}
{\large 
Hyewon Han\footnote{dwhw101@dgu.ac.kr}, Bogeun Gwak\footnote{rasenis@dgu.ac.kr}
} \\
\vspace*{0.5cm}

{\it 
Department of Physics, Dongguk University, Seoul 04620, Republic of Korea
}

\end{center}

\vspace*{1.0cm}
\begin{abstract}
We study the correspondence between quasinormal modes (QNMs) and grey-body factors (GBFs) of higher-dimensional Reissner–Nordstr\"{o}m black holes. Considering scalar, vector, and tensor gravitational perturbations, we systematically compute the QNMs and GBFs for each spacetime dimension and set of black-hole parameters to evaluate the analytical formula connecting them. QNMs are obtained by the continued fraction and the integration-through-midpoints methods depending on the singularity structure in the Frobenius series. We compare the GBFs determined by the correspondence using QNMs with those computed numerically. The correspondence performs well for scalar($+$), vector($+$), vector($-$), and tensor perturbations. For the scalar($-$) perturbation, the GBFs obtained by the correspondence deviate noticeably from the accurate GBFs for black holes with low to intermediate values of the charge in $D>6$. The correspondence recovers reasonable accuracy at higher charges, although the effective potential exhibits double peaks, contrary to the assumption underlying the Wentzel–Kramers–Brillouin (WKB)-based derivation of the correspondence. Whereas the accuracy decreases with increasing $D$ for other perturbation types, the correspondence for the scalar($-$) perturbation yields more accurate GBFs in $D=8$ than those in $D=7$. For near-extreme black holes, the correspondence offers a reliable approximation of GBFs for the vector($-$) perturbation, even though a double-peak potential occurs in $D>6$. These results suggest that the correspondence has a wider range of applicability than might be expected from the assumption of the WKB approach.  
\end{abstract}

\newpage
\baselineskip=18pt
\setcounter{page}{2}
\pagestyle{plain}
\baselineskip=18pt
\pagestyle{plain}
\setcounter{footnote}{0}

\hrule
\vspace{1em}

\tableofcontents

\vspace{3em}
\hrule

\vspace{1em}

\section{Introduction}
A black hole is an important system for studying physics in strong gravitational fields. It is enclosed by an event horizon that serves as a causal boundary between its interior and exterior. The causal structure inside the event horizon prevents infalling matter from escaping to the exterior and constrains its motion toward the curvature singularity. The spacetime geometry and the horizon structure of a black hole are determined by parameters such as its mass, angular momentum, and electric charge. When all three parameters are nonzero, a rotating and charged black hole can be described by the Kerr–Newman solution. In the non-rotating limit, this solution reduces to the Reissner–Nordstr\"{o}m solution, whose neutral limit yields the Schwarzschild solution. For a rotating black hole, the horizon is surrounded by an ergosphere from which the energy of the black hole can be extracted \cite{Penrose:1969pc}. Although the black hole’s total mass decreases through energy extraction, its irreducible mass cannot decrease in classical processes \cite{Christodoulou:1970wf,Christodoulou:1971pcn}. This property, analogous to the nondecreasing nature of entropy in the second law of thermodynamics, motivated the introduction of black-hole entropy. Black holes act as thermodynamic systems, with the Bekenstein–Hawking entropy proportional to the area of the event horizon \cite{bekenstein2020black,hawking1975particle}. Furthermore, quantum effects near the horizon give rise to thermal radiation, known as Hawking radiation, characterized by a temperature related to the surface gravity at the horizon \cite{hawking1975particle}. These findings reveal an intriguing connection between classical black-hole dynamics and quantum physics. 

Gravity theories can be extended to spacetime with more than four dimensions. Higher-dimensional gravity has been motivated by various theoretical developments, including string theory and the brane-world scenario \cite{Arkani-Hamed:1998jmv,Randall:1999vf}. The Schwarzschild solution generalizes to the Schwarzschild–Tangherlini solution \cite{tangherlini1963schwarzschild} in higher dimensions, while charged and rotating black holes are described by the higher-dimensional Reissner–Nordstr\"{o}m and Myers–Perry solutions \cite{Myers:1986un,Myers:2011yc}, respectively. The number of spacetime dimensions affects the geometry and dynamical properties of black holes, allowing for features that have no counterpart in four-dimensional gravity. Myers–Perry black holes possess multiple angular momenta related to distinct rotation planes and can suffer from ultraspinning instabilities \cite{Emparan:2003sy,Dias:2010maa}. Moreover, higher-dimensional gravity allows for the existence of other black objects, such as black strings and black branes, which exhibit horizon geometries and topologies distinct from those of spherical black holes and can be subject to the Gregory–Laflamme instability \cite{Gregory:1993vy,Gregory:1994bj,Harmark:2007md,Dias:2010eu}. 

The dynamical behavior and stability of black holes can be studied by examining their response to perturbations. When a perturbed black hole settles down to a stationary state, its response exhibits damped oscillations known as quasinormal modes (QNMs) \cite{Berti:2009kk,Konoplya:2011qq,Zhidenko:2006rs,Konoplya:2007jv,Konoplya:2008au,Ponglertsakul:2020ufm,Gwak:2022nsi,BarraganAmado:2023wxt,Lin:2024ubg,Han:2024rus}. Their complex frequencies are called quasinormal (QN) frequencies and are determined by the boundary conditions under which waves move purely inward at the event horizon and purely outward to spatial infinity. The real component of a QN frequency specifies the oscillation frequency, and its imaginary component determines the decay rate of the perturbation. The QNMs are labeled by an overtone index, and the fundamental mode, with the smallest magnitude of the imaginary component, corresponds to the least damped mode. Because QN frequencies depend on the parameters of black holes, they carry characteristic information about the background spacetime \cite{Dreyer:2003bv}. 
QNMs are relevant to the analysis of instabilities in black-hole spacetime. Under suitable conditions on the frequency, waves propagating around a rotating black hole can extract rotational energy and undergo amplification called superradiance \cite{Press:1972zz,Cardoso:2004nk}. For perturbations of massive bosonic fields or under the boundary conditions of asymptotically anti-de Sitter (AdS) black holes with a negative cosmological constant, an effective confining mechanism causes the waves to undergo repeated amplification. This process leads to a strong instability known as the black-hole bomb \cite{Brito:2015oca}. Small Kerr-AdS black holes are unstable under scalar and gravitational perturbations through this mechanism \cite{Cardoso:2004hs,Cardoso:2006wa}. Superradiance has also been identified in Myers–Perry-AdS black holes \cite{Kunduri:2006qa,Kodama:2009rq}. For massive, charged scalar field perturbations, Reissner–Nordstr\"{o}m black holes can suffer from superradiant instabilities in four and higher dimensions \cite{Herdeiro:2013pia,Hod:2013fvl,Zhu:2014sya,Destounis:2019hca}.
The decay rate of QNMs is related to the stability of the inner Cauchy horizon. Waves propagating toward the Cauchy horizon experience a blueshift, and the stability of the horizon depends on the competition between its surface gravity and the decay rates of perturbations \cite{Cardoso:2017soq}. This competition is central to investigations of strong cosmic censorship \cite{Penrose:1969pc}, which asserts the inextendibility of spacetime beyond the Cauchy horizon. Strong cosmic censorship has been tested for scalar perturbations in Reissner–Nordstr\"{o}m-de Sitter (dS) and Kerr-dS black holes \cite{Cardoso:2017soq,Dias:2018ynt}, and subsequent studies have considered other field perturbations \cite{Destounis:2018qnb,Mo:2018nnu,Gwak:2018rba,Gim:2019rkl} and higher-dimensional spacetime \cite{Liu:2019lon,Liu:2019rbq}.

Scattering of waves in black-hole spacetime is also important for exploring its physical properties. When an incident wave encounters the effective potential barrier, a fraction of the wave is reflected, while the remainder is transmitted through the barrier. The boundary conditions of this scattering process allow for both incoming and reflected waves at infinity but only ingoing waves at the event horizon. The probability that the incident wave penetrates the potential is referred to as the grey-body factor (GBF) \cite{hawking1975particle,Page:1976df,Cornell:2005ux,Cardoso:2005vb,Creek:2006ia,Harmark:2007jy}. Due to the symmetry of the process, the same probability can be obtained for the radiation emitted near the event horizon. Hence, the GBF quantifies the deviation of the black-hole radiation from that of an ideal blackbody. It is essential for evaluating the Hawking emission spectrum and the evaporation rate of black holes \cite{hawking1975particle,Page:1976df}. Studies have shown that GBFs can be used to model the spectral amplitudes of ringdown signals emitted by black-hole mergers \cite{Oshita:2022pkc,Oshita:2023cjz,Okabayashi:2024qbz}. As the GBF depends on the properties of the final remnant, it facilitates the estimation of remnant parameters and the test of the no-hair theorem.

QNMs and GBFs, although characterized by distinct boundary conditions, can be analytically connected. A correspondence between them was originally constructed for spherically symmetric black holes in four dimensions \cite{Konoplya:2024lir} using the Wentzel–Kramers–Brillouin (WKB) approximation \cite{Schutz:1985km,Iyer:1986np,Konoplya:2019hlu}. The correspondence provides exact GBFs from QN frequencies in the eikonal limit, characterized by a high multipole $l$, but only approximate ones away from this limit. Therefore, its precision at finite $l$ requires validation. The correspondence has been applied to various black holes, including rotating black holes \cite{Konoplya:2024vuj,Pedrotti:2025idg}, dS black holes \cite{Malik:2024cgb}, regular black holes \cite{Heidari:2024bbd,Tang:2025mkk,Lutfuoglu:2025ohb,Shi:2025gst,Bolokhov:2025lnt,Ji:2025nlc,Malik:2025dxn,Dubinsky:2025nxv,Lutfuoglu:2025blw,Malik:2025qnr,Dubinsky:2025wns,Lutfuoglu:2025mqa,Dubinsky:2026wcv,Bolokhov:2026eqf,Bolokhov:2026kqu,Lutfuoglu:2026rqe}, and black holes in modified gravity theories \cite{AraujoFilho:2025hkm,Heidari:2025oop,Konoplya:2025mvj,Lutfuoglu:2025hjy,Sajadi:2025kah,Lutfuoglu:2025ldc,Hamil:2025fbn,Bolokhov:2026uol}. It was first generalized to five dimensions in \cite{Han:2025cal} and subsequently to higher dimensions in \cite{Han:2026fpn}. For Schwarzschild–Tangherlini black holes, the correspondence for scalar gravitational perturbations was found to fail when the effective potential departs from the structure for which the standard WKB approach can be applied. Furthermore, the correspondence breaks down when the potential is not characterized by a centrifugal form in the eikonal regime, as in black holes with higher-curvature corrections \cite{Konoplya:2017wot,Konoplya:2025afm}. The correspondence has also been extended to scalar and electromagnetic perturbations of higher-dimensional regular black holes \cite{Arbelaez:2026eaz,Arbelaez:2026cnd,Fan:2026mdu}.

In this study, the correspondence is generalized to charged black holes in higher dimensions. We employ the higher-dimensional Reissner–Nordstr\"{o}m metric and consider its gravitational perturbations, consisting of scalar, vector, and tensor types \cite{Kodama:2003jz,Kodama:2003kk,Cardoso:2003vt,Lunin:2025yth}. The first two types are further classified into ($+$) and ($-$) types due to the coupling of gravitational and electromagnetic perturbations. Unlike the four-dimensional black hole, for which the scalar and vector types lead to the same QNM spectrum, the higher-dimensional black hole enables us to analyze the correspondence for the distinct spectra of all perturbation types, including the tensor type, which appears only in higher dimensions. To obtain the GBFs through the correspondence, we calculate accurate QNMs using a continued fraction method \cite{Leaver:1985ax,Leaver:1990zz}. The original setting of this method cannot be applied when the singularities of the wave equation, which depend on perturbation types and spacetime dimension, hinder the convergence of the Frobenius series. For these cases, an integration-through-midpoints method is used \cite{Rostworowski:2006bp}. We present our results for the fundamental QNM and the first overtone; quantitative data for the latter have been limited. The correspondence yields approximate GBFs at low multipoles. We compare them with the accurate GBFs obtained by numerical computation, and the discrepancy between the two quantifies the performance of the correspondence. We systematically evaluate the accuracy of the approximate GBFs for all types of perturbation across spacetime dimensions and black-hole charges and discuss the applicability of the correspondence.

We organize the remainder of the paper as follows. Section $2$ presents the background metric and the wave equation for its gravitational perturbations. Section $3$ introduces the analytical relation between QNMs and GBFs. In Section $4$, the QN frequencies are calculated using the continued fraction and the integration-through-midpoints methods, and the GBFs are numerically computed to evaluate the correspondence. In Sections $5,6,$ and $7$, the GBFs are obtained from the correspondence for the scalar, vector, and tensor types of perturbation, respectively. Finally, Section $8$ summarizes the results. 

We set $c=G_D=1$ and adopt the signature $(-, +, +, +, \cdots)$ throughout this paper.

\section{Perturbation equation of higher-dimensional Reissner–Nordstr\"{o}m black holes}
We study the relationship between the QNMs and the GBFs by considering gravitational perturbations of spherically symmetric, charged, and asymptotically flat black holes in higher dimensions. The corresponding spacetime can be described by the higher-dimensional Reissner–Nordstr\"{o}m solution. In this section, the background metric and the equation governing its gravitational perturbations are briefly reviewed. 

The higher-dimensional Reissner–Nordstr\"{o}m black hole is a solution to the Einstein–Maxwell field equations obtained from the action
    \begin{align}
        S =\frac{1}{16\pi} \int d^D x \sqrt{-g} \, \left(R - F_{\alpha \beta} F^{\alpha \beta} \right),
    \end{align}
where $D>3$ is the spacetime dimension, $g$ the metric determinant, $R$ the Ricci scalar, and $F^{\alpha \beta}$ the electromagnetic field-strength tensor. The metric is given by
    \begin{align} \label{metric}
        ds^2=-f(r) dt^2 +\frac{1}{f(r)}dr^2+r^2 d \Omega_{\mu}^2 ,
    \end{align}
where $d \Omega_{\mu}^2$ is the line element on the unit $\mu$-sphere with $\mu=D-2$. We express the metric function as
    \begin{align} \label{metricfunction}
        f(r)=1-\frac{2M}{r^{\mu-1}}+\frac{Q^2}{r^{2\mu-2}}.
    \end{align}
Here, the mass parameter $M$ is proportional to the Arnowitt–Deser–Misner (ADM) mass $M_B$ as
    \begin{align}
        M= \frac{8 \pi M_B}{\mu \Omega_{\mu}},
    \end{align}
with $\Omega_\mu=2\pi^{\frac{\mu+1}{2}}/\Gamma(\frac{\mu+1}{2})$. The total charge $Q_B$ of the black hole is related to the charge parameter $Q$ as
    \begin{align}
        Q = \sqrt{\frac{2}{\mu (\mu-1)}}\frac{4 \pi Q_B}{\Omega_{\mu}}.
    \end{align}
The metric \eqref{metric} recovers the Schwarzschild–Tangherlini metric \cite{tangherlini1963schwarzschild} in the limit $Q \to 0$. Considering nonzero $Q$, we find the horizons of the black hole at
    \begin{align} \label{horizons}
        r_{\pm}=\left(M \pm \sqrt{M^2 -Q^2} \right)^{1/(\mu-1)}.
    \end{align}
The outer horizon $r_+$ corresponds to the event horizon, and we denote it by $r_H$. As can be seen from Eq. \eqref{horizons}, two horizons exist only when $|Q| < M$. When the charge and mass parameters are equal in magnitude, $|Q|=M$, the horizons coincide, resulting in the extreme black hole. We analyze non-extreme cases with $|Q|<M$ in this work.

We consider gravitational perturbations in the fixed background metric \eqref{metric}. Gravitational perturbations of higher-dimensional Reissner–Nordstr\"{o}m black holes are categorized into scalar, vector, and tensor types \cite{Kodama:2003jz,Kodama:2003kk}. The first two types reduce to polar and axial perturbations in four-dimensional black holes, respectively. Unlike the four-dimensional black hole, for which the two types have an identical QN spectrum, the higher-dimensional black hole allows us to study the distinct spectra for the scalar and vector types. Because gravitational perturbations of charged black holes are coupled with electromagnetic perturbations, the scalar and vector perturbations are each divided into ($+$) and ($-$) types. In the limit of $Q \to 0$, scalar($+$) and vector($+$) gravitational perturbations reduce to perturbations of the test electromagnetic fields, while scalar($-$) and vector($-$) perturbations recover the scalar and vector gravitational perturbations of neutral black holes. Tensor gravitational perturbations, present only in higher dimensions, also have a QN spectrum distinct from those of the other types. We study the relationship between QNMs and GBFs for all gravitational perturbations.

A master equation of gravitational perturbations in higher-dimensional Reissner–Nordstr\"{o}m black holes is given by the Schr\"{o}dinger-type equation \cite{Kodama:2003kk}
    \begin{align} \label{waveeq}
        \frac{d^2 \Psi}{dr_*^2}+\left( \omega^2 - V(r) \right) \Psi=0,
    \end{align}
where $r_*=\int \frac{dr}{f}$ is the tortoise coordinate, $\omega$ is the frequency of the wave, and $V(r)$ is an effective potential. The scalar types of perturbation are described by the potentials
    \begin{align} \label{VS}
        V_{S \pm}(r)&= \frac{f U_{\pm}}{64 r^2 H_{\pm}^2} ,
    \end{align}
where 
    \begin{align}
        H_{+}&=1-\frac{\mu (\mu+1)}{2} \delta x, \\
        H_{-}&=m+\frac{\mu (\mu+1)}{2} (1+m\delta) x, \\
        U_+ &= -\delta^3 \mu^3(3 \mu-2)(\mu+1)^4 (1+m\delta)x^4 \nonumber  \\
        & \quad \,+4 \delta^2 \mu^2 (\mu+1)^2 \left\{(\mu+1)(3\mu-2)m\delta+4\mu^2+\mu-2\right\}x^3 \nonumber \\
        & \quad \,+4 \delta(\mu+1) \left\{(\mu-2)(\mu-4)(\mu+1)(m+\mu^2)\delta-7 \mu^3+7\mu^2-14\mu+8 \right\} x^2 \nonumber \\
        & \quad \,+\left\{16(\mu+1)(-4m+3\mu^2(\mu-2))\delta-16(3\mu-2)(\mu-2)\right\}x  \nonumber \\
        & \quad \,+64m+16\mu(\mu+2),   \\
        U_- &= -\mu^3(3 \mu-2)(\mu+1)^4 \delta (1+m\delta)^3 x^4 \nonumber  \\
        & \quad \,-4 \mu^2 (\mu+1)^2 (1+m\delta)^2 \left\{(\mu+1)(3\mu-2)m\delta-\mu^2 \right\}x^3 \nonumber \\
        & \quad \,+4 (\mu+1)(1+m\delta) \left\{m(\mu-2)(\mu-4)(\mu+1)(m+\mu^2)\delta \right. \nonumber \\
        & \left. \quad \quad +4\mu (2 \mu^2-3\mu+4)m+\mu^2(\mu-2)(\mu-4)(\mu+1) \right\} x^2 \nonumber \\
        & \quad \,-16 m\left\{(\mu+1)m(-4m+3\mu^2(\mu-2))\delta +3\mu (\mu-4)m +3\mu^2 (\mu+1)(\mu-2)\right\}x \nonumber \\
        & \quad \,+64m^3+16\mu(\mu+2)m^2,   \\
        m&=l(l+\mu-1)-\mu, \\
        \delta &=\frac{1}{2m}\left(\sqrt{1+\frac{4mQ^2}{(\mu+1)^2 M^2}}-1 \right), \\
        x&=\frac{2M}{r^{\mu-1}}, \\
        z&=\frac{Q^2}{r^{2\mu-2}},
    \end{align}
with a multipole number $l$. The plus and minus indices correspond to the scalar($+$) and scalar($-$) types, respectively. Vector and tensor perturbations are governed by the equation \eqref{waveeq} with effective potentials
    \begin{align}
        V_{V\pm}(r)&=\frac{f}{r^2}\left[ l (l + \mu - 1) - 1 + \frac{\mu^2 - 2 \mu + 4}{4} + \frac{\mu (5 \mu - 2)}{4} z - \frac{\mu^2 + 2}{4} x \right. \nonumber \\
        & \quad \quad \left. \pm 
   \frac{1}{r^{\mu - 1}} \sqrt{(\mu^2 - 1)^2 M^2 + 2\mu (\mu - 1) m Q^2}\right], \label{VV} \\
        V_{T}(r)&=\frac{f}{r^2}\left[ l (l + \mu - 1) -2 + 2\mu + \frac{\mu^2 - 10 \mu + 8}{4} + \frac{\mu (3 \mu - 2)}{4} z - \frac{\mu^2}{4} x \right], \label{VT}
    \end{align}
respectively. Similarly, the plus and minus indices in $V_{V\pm}$ indicate the vector($+$) and vector($-$) types, respectively.

For convenience, we introduce dimensionless quantities as 
    \begin{align} \label{dimless}
        \hat{t}=\frac{t}{r_H}, \qquad \hat{r}=\frac{r}{r_H}, \qquad \hat{M}=\frac{M}{r_H^{\mu-1}}, \qquad \hat{Q}=\frac{Q}{r_H^{\mu-1}}, \qquad \hat{\omega}=r_H \omega.
    \end{align} 
Then, we rewrite the metric function \eqref{metricfunction} as
    \begin{align}
        f(\hat{r})= 1-\frac{2 \hat{M}}{\hat{r}^{\mu-1}}+\frac{\hat{Q}^2}{\hat{r}^{2 \mu -2}}.
    \end{align}
In this notation, the event horizon lies at $\hat{r}=1$, which gives the relation
    \begin{align}
        2 \hat{M}=1+\hat{Q}^2.
    \end{align}
The condition for the extreme black hole $|\hat{Q}| = \hat{M}$ gives the extreme value of the charge parameter, $\hat{Q}_{ext}=1$. We thus consider the non-extreme black holes with $\hat{Q}<1$. We henceforth omit the hats in Eq. \eqref{dimless} and use the same symbols for notational simplicity.

\section{Correspondence between QNMs and GBFs}
QNMs and GBFs are distinct spectral quantities that characterize the scattering and radiative properties around the black holes. They are determined by solving the wave equation \eqref{waveeq} with their respective boundary conditions. As shown in \cite{Konoplya:2024lir}, the GBFs can be obtained from QN frequencies through a correspondence based on the WKB approach. In this section, we introduce the analytical formula connecting QNMs and GBFs. 

A QNM is a damped oscillation mode satisfying a purely ingoing boundary condition at the event horizon and a purely outgoing condition at spatial infinity. Its associated QN frequency is complex, $\omega=\omega_R+i \omega_I$, where $\omega_R$ and $\omega_I$ determine the oscillation frequency and the decay rate of the wave, respectively. When we assume the time dependence of the wave to be $e^{-i\omega t}$ and the real part to be positive, $\omega_R>0$, the asymptotic solutions are written as
    \begin{align} \label{qnmbc}
        \Psi=\begin{cases}
                    \, e^{+ i \omega r_*}, & \quad\mbox{for } \,r_* = + \infty, \\
                    \, e^{- i \omega r_*} , & \quad \mbox{for } \, r_* = - \infty,
            \end{cases}
    \end{align}
where the radial tortoise coordinate $r_*=\pm\infty$ corresponds to spatial infinity and the event horizon, respectively. By solving Eq. \eqref{waveeq} under these boundary conditions, one obtains the QN frequencies for a given set of black-hole parameters and multipole number.

The GBFs are associated with the scattering problem near the black hole. If we assume waves enter from infinity, they partially penetrate the effective potential barrier of the black hole and are partially reflected by the barrier. Then, the asymptotic solutions at the boundaries are given by
    \begin{align} \label{scatteringbc}
        \Psi=\begin{cases}
                    \, e^{- i \Omega r_*} + R  \, e^{+i \Omega r_*}, & \quad\mbox{for } \,r_* = + \infty, \\
                    \, Te^{- i \Omega r_*} , & \quad \mbox{for } \, r_* = - \infty,
                \end{cases}
        \end{align} 
where $R$ is the reflection coefficient and $T$ is the transmission coefficient. Here, a real frequency $\Omega$ is distinguished from the complex QN frequency $\omega$. The GBF is defined as the transmission probability through the potential,
    \begin{align} 
        \Gamma(\Omega) = |T|^2 = 1- |R|^2.
    \end{align}
Because the scattering process in black-hole spacetime is symmetric, the same transmission probability can be obtained for the waves emitted near the event horizon. Thus, the GBF measures the deviation of the radiation of black holes from blackbody radiation. 

GBFs are usually computed by numerically solving the wave equation \eqref{waveeq}. However, they can be determined through an analytical formula in terms of the QN frequencies. In \cite{Konoplya:2024lir}, the correspondence between QNMs and GBFs was established using the higher-order WKB approach. In spacetimes where the WKB approach exactly describes the eikonal regime, the correspondence provides exact GBFs using the fundamental QN frequency. For low multipoles, it yields only approximate GBFs because of the asymptotic properties of WKB series beyond the eikonal limit.

For an effective potential that possesses a single barrier with two turning points and vanishes asymptotically, the WKB approach expands the potential around its peak and matches the solution to the asymptotic WKB solution. The standard WKB formula for the QN frequency $\omega$ is given by 
    \begin{align} \label{wkb}
        i \frac{\omega^2-V_0}{\sqrt{-2V_0''}}-\sum^{k}_{i=2}\Lambda_i=\mathcal{K},
    \end{align}
where $V_0$ is the maximum value of the effective potential, $V_0 ''$ is the second derivative of the potential with respect to $r_*$ at its maximum, and $\Lambda_i$ denotes the $i$th-order correction, which can be found up to the $16$th order in \cite{Iyer:1986np,Konoplya:2003ii,Matyjasek:2017psv,Matyjasek:2019eeu,Matyjasek:2026yiu}. For the QN boundary conditions \eqref{qnmbc}, $\mathcal{K}=n+\frac{1}{2}$ with an overtone index $n=0,1,2, \cdots$. We represent the frequency corresponding to the overtone $n$ as $\omega_n$. The least damped mode $\omega_0$, with the smallest magnitude of the imaginary component, is the fundamental mode. 

The WKB expression \eqref{wkb} can be written for the scattering problem \eqref{scatteringbc}, with $\omega$ replaced by a real frequency $\Omega$. Then, the GBF is expressed as
    \begin{align} \label{gbf}
        \Gamma (\Omega)=\frac{1}{1+e^{2 \pi i \mathcal{K}}},
    \end{align}
where $\mathcal{K}$ is determined from the WKB formula for each $\Omega$. 

Incorporating the sixth-order WKB formula for the QNMs and for the scattering process yields an analytical relation between QN frequencies and GBFs as \cite{Konoplya:2024lir}
    \begin{align}
         i\mathcal{K}=& \frac{\Omega^2 - \mathrm{Re}[\omega_0]^2}{4 \, \mathrm{Re}[\omega_0] \, \mathrm{Im}[\omega_0]} - \frac{\mathrm{Re}[\omega_0]-\mathrm{Re}[\omega_1]}{16 \, \mathrm{Im}[\omega_0]} \nonumber \\
        & +\frac{\Omega^2 - \mathrm{Re}[\omega_0]^2}{32 \, \mathrm{Re}[\omega_0] \, \mathrm{Im}[\omega_0]} \left(\frac{\left(\mathrm{Re}[\omega_0]-\mathrm{Re}[\omega_1]\right)^2}{4 \, \mathrm{Im}[\omega_0]^2} - \frac{3\mathrm{Im}[\omega_0]-\mathrm{Im}[\omega_1]}{3 \, \mathrm{Im}[\omega_0]} \right) \nonumber  \\
        &- \frac{\left(\Omega^2 - \mathrm{Re}[\omega_0]^2\right)^2}{16 \, \mathrm{Re}[\omega_0]^3 \, \mathrm{Im}[\omega_0]} \left(1+\frac{\mathrm{Re}[\omega_0]\left(\mathrm{Re}[\omega_0]-\mathrm{Re}[\omega_1]\right)}{4 \, \mathrm{Im}[\omega_0]^2} \right) \nonumber \\
        &+\frac{\left(\Omega^2 - \mathrm{Re}[\omega_0]^2\right)^3}{32 \, \mathrm{Re}[\omega_0]^5 \, \mathrm{Im}[\omega_0]} \left(1+\frac{\mathrm{Re}[\omega_0]\left(\mathrm{Re}[\omega_0]-\mathrm{Re}[\omega_1]\right)}{4 \, \mathrm{Im}[\omega_0]^2} \right. \nonumber \\
        & \left. \qquad \qquad \qquad + \mathrm{Re}[\omega_0]^2 \left( \frac{\left(\mathrm{Re}[\omega_0]-\mathrm{Re}[\omega_1]\right)^2}{16 \, \mathrm{Im}[\omega_0]^4} - \frac{3\mathrm{Im}[\omega_0]-\mathrm{Im}[\omega_1]}{12 \, \mathrm{Im}[\omega_0]^3} \right)\right) + \mathcal{O}(l^{-3}).\label{corr}
    \end{align}
The GBFs can be calculated by substituting the values of the fundamental QN frequency $\omega_0$ and the first overtone $\omega_1$ into this relation. When the effective potential lies within the class where the WKB method is applicable, the exact GBFs can be obtained in the eikonal limit through the leading term on the right side of Eq. \eqref{corr}, a term that depends only on the fundamental mode. Beyond the eikonal limit, the remaining correction terms, which also depend on the first overtone, improve the accuracy of the GBF. 

Since the correspondence yields only approximate GBFs at low multipoles, its validity is generally not guaranteed. 
The effective potentials of gravitational perturbations in the higher-dimensional Reissner–Nordstr\"{o}m black hole have a centrifugal term $ \sim f \frac{l(l+1)}{r^2}$ in the eikonal regime, in which the WKB approximation works well. The correspondence is therefore also expected to perform well in this regime. In this work, we investigate its applicability at a low multipole $l=2$.

\section{Numerical computation of QNMs and GBFs}
To obtain the GBFs from the correspondence \eqref{corr}, we compute the fundamental QNM $\omega_0$ and the first overtone $\omega_1$. We employ the continued fraction method \cite{Leaver:1985ax} to find accurate QN frequencies. It derives a recurrence relation for the coefficients of the Frobenius series and constructs infinite continued fractions from the convergence conditions of the series. The original setting of this technique fails when the Frobenius series does not converge between the event horizon and spatial infinity. In that case, we adopt the modified method developed in \cite{Rostworowski:2006bp}, which introduces midpoints to ensure that the series converges in this region. For higher-dimensional Reissner–Nordstr\"{o}m black holes, the singularity structure in the Frobenius series depends on the type of perturbation, the spacetime dimension, and the charge of the black hole. We calculate the QNMs using the appropriate method for each case.

We start by introducing a new function
    \begin{align} 
        \Phi=e^{-i \omega r_*} \Psi.
    \end{align}
The wave equation \eqref{waveeq} is then expressed as
    \begin{align} \label{waveeqtr}
        \frac{d^2 \Phi}{dr_*^2}+2 i \omega \frac{d \Phi}{dr_*}- V \Phi=0.
    \end{align}
The ansatz for the wave function satisfying the QN boundary conditions \eqref{qnmbc} can be written as \cite{Konoplya:2007jv}
    \begin{align} \label{ansatz}
        \Phi=\left(1-\frac{1}{r}\right)^{-2 i \omega/f'(1)} \sum^{\infty}_{k=0} a_k \left( \frac{r-1}{r-r_-} \right)^k,
    \end{align}
where $f'(r)$ is the derivative of the metric function with respect to $r$. In the present notation, the event horizon is located at $r=1$, and the inner horizon at $r_-=(Q^2)^{\frac{1}{\mu-1}}$. Upon substituting Eq. \eqref{ansatz} into Eq. \eqref{waveeqtr} with the effective potentials for each perturbation type in Eqs. \eqref{VS}, \eqref{VV}, and \eqref{VT}, we derive the $N$-term recurrence relation,
    \begin{align} \label{recur}
        \sum^{\min(N-1,k+1)}_{j=0}b^{(N)}_{j,k}  \, a_{k+1-j}=0, \qquad \mathrm{for} \ k=0,1,2,\cdots,
    \end{align}
where the coefficients $\{b^{(N)}_{j,k}\}$ are functions of the QN frequency $\omega$ for fixed values of the black-hole parameters and the multipole number $l$. The number $N$ of terms in Eq. \eqref{recur} is determined by the spacetime dimension and the perturbation type. To construct the continued fraction for the coefficients $\{b^{(N)}_{j,k}\}$, we reduce the $N$-term recurrence relation to a three-term relation through the Gaussian elimination \cite{Zhidenko:2006rs,Konoplya:2011qq} as
    \begin{align} \label{GaE}
        b^{(p)}_{j,k}&=b^{(p+1)}_{j,k} \qquad \mathrm{for} \ j=0, \ \mathrm{or} \ (k+1) < p, \nonumber \\
        b^{(p)}_{j,k}&=b^{(p+1)}_{j,k}-\frac{b^{(p+1)}_{p,k} \, b^{(p)}_{j-1,k-1}}{b^{(p)}_{p-1,k-1}}.
    \end{align}
Then, by defining $\alpha_k=b^{(3)}_{0, k}, \beta_k=b^{(3)}_{1, k}$, and $\gamma_k=b^{(3)}_{2, k}$, we find the three-term recurrence relation as
    \begin{align} 
        \alpha_0 & a_1+\beta_0 a_0 =0,\\
        \alpha_k & a_{k+1}+\beta_k a_k + \gamma_k a_{k-1}=0, \qquad \mathrm{for} \ k>0, \label{threerc}
    \end{align}
where we assume $a_0=1$. From the convergence condition of the series in Eq. \eqref{ansatz}, we obtain the continued fraction as
    \begin{align} \label{n0}
        0=\beta_0 - \frac{\alpha_0 \gamma_1}{\beta_1-\frac{\alpha_1 \gamma_2}{\beta_2-\frac{\alpha_2 \gamma_3}{\beta_3- \cdots}}} \equiv \beta_0 - \frac{\alpha_0 \gamma_1}{\beta_1 -}\frac{\alpha_1 \gamma_2}{\beta_2 -}\frac{\alpha_2 \gamma_3}{\beta_3 -} \cdots.
    \end{align}
The fundamental mode $\omega_0$ is obtained by numerically solving this equation, and higher overtones are calculated from the inverted equation
    \begin{align} 
        \beta_n-\frac{\alpha_{n-1}\gamma_n}{\beta_{n-1} -}\frac{\alpha_{n-2}\gamma_{n-1}}{\beta_{n-2} -}\cdots\frac{\alpha_0 \gamma_1}{\beta_0}=\frac{\alpha_n \gamma_{n+1}}{\beta_{n+1} -}\frac{\alpha_{n+1} \gamma_{n+2}}{\beta_{n+2} -}\frac{\alpha_{n+2} \gamma_{n+3}}{\beta_{n+3} -} \cdots, \quad \mathrm{for}\ n \ge1 .
    \end{align}
To accelerate the convergence of the continued fractions, we use the Nollert improvement \cite{Nollert:1993zz}, which replaces the tail of the truncated continued fraction with an approximate value. 

The continued fraction method is not applicable if any singularity of the wave equation occurs within a unit sphere centered at $\rho \equiv \frac{r-1}{r-r_-}=0$. These singularities prevent the series from converging at $\rho=1$, which corresponds to spatial infinity. In this case, we use the integration-through-midpoints method \cite{Rostworowski:2006bp,Konoplya:2011qq} which introduces regular points inside the radius of convergence of the initial series and expands the series at these points until $\rho=1$ lies within the radius of convergence. By choosing a midpoint $\rho_0$, we have
    \begin{align} \label{news}
        \sum^{\infty}_{k=0} a_k \rho^k=\sum^{\infty}_{k=0} a'_k (\rho-\rho_0)^k,
    \end{align}
where
    \begin{align} 
        a'_0=\sum^{\infty}_{k=0} a_k \rho_0^k, \qquad a'_1=\sum^{\infty}_{k=1} k a_k \rho_0^{k-1}.
    \end{align}
The coefficients $\{a'_k\}$ satisfy a new recurrence relation, whose number of terms can also be reduced to three as in Eq. \eqref{threerc} using Gaussian elimination. Consequently, we construct the continued fraction equation that yields the QN frequency. If $\rho=1$ is not the closest singular point to the midpoint $\rho=\rho_0$, we introduce further midpoints $\rho_1,\rho_2, \cdots$ to ensure that the radius of convergence covers $\rho=1$.

Furthermore, we compute the GBFs using a numerical method to evaluate the accuracy of the GBFs obtained by the relationship with QNMs. To solve the wave equation
    \begin{align} \label{waveeq2}
        \frac{d^2 \Psi}{dr_*^2}+\left( \Omega^2 - V \right) \Psi=0,
    \end{align}
and determine the GBFs for a given frequency $\Omega$, we employ the Mathematica-based tool \texttt{GrayHawk} \cite{Calza:2025whq,Calza:2026wuf}. It is publicly available for computing the GBFs of various black holes and wormholes, and we apply it to the higher-dimensional Reissner–Nordstr\"{o}m black holes. The numerically calculated GBFs are used as reference values for comparison with the GBFs obtained from the correspondence \eqref{corr}.

\section{Correspondence in scalar-type perturbations}
We analyze the correspondence between QNMs and GBFs for scalar gravitational perturbations across the number of spacetime dimensions $D>4$ and the charge parameter $Q$ of black holes. To examine the effects of charge, we calculated QNMs and GBFs for $Q$ ranging from $0$ to $0.9$ and also obtained results for near-extreme black holes with $Q=0.98$.

\subsection{Scalar($+$) perturbation}
The scalar($+$) type of gravitational perturbation is described by the effective potential $V_{S+}$ \eqref{VS}. To obtain the GBFs using the correspondence, we calculated the fundamental mode and the first overtone. The results for the five-dimensional black holes are shown in \autoref{qnmS+5}. As the charge parameter $Q$ grows, the real part of the frequency first increases and subsequently decreases, whereas the magnitude of the imaginary part monotonically decreases. 

\begin{table}[h!]
\centering
\begin{tabular}{
|>{\centering\arraybackslash}p{1.5cm}|
 >{\centering\arraybackslash}p{4.5cm}|
 >{\centering\arraybackslash}p{4.5cm}| }

\hline\hline
\multicolumn{3}{|c|}{\textbf{$D=5, \, l=2$}} \\
\hline\hline

$Q$ & $n=0$ & $n=1$ \\
\hline
$0$   & $1.341207 - 0.337563 i$ & $1.212023 - 1.046187 i$ \\
$0.1$   & $1.346208 - 0.336826 i$ & $1.218905 - 1.043459 i$ \\
$0.2$   & $1.359563 - 0.334198 i$ & $1.237904 - 1.034059 i$ \\
$0.3$   & $1.377265 - 0.328710 i$ & $1.264821 - 1.015143 i$ \\
$0.4$   & $1.394543 - 0.319371 i$ & $1.294131 - 0.983774 i$ \\
$0.5$   & $1.406690 - 0.305577 i$ & $1.319513 - 0.938166 i$ \\
$0.6$   & $1.409221 - 0.287425 i$ & $1.333879 - 0.878760 i$ \\
$0.7$   & $1.398017 - 0.266102 i$ & $1.329787 - 0.809781 i$ \\
$0.8$   & $1.370076 - 0.244259 i$ & $1.302273 - 0.741066 i$ \\
$0.9$   & $1.325187 - 0.225521 i$ & $1.254967 - 0.684736 i$ \\
\hline\hline

\end{tabular}
    \caption{QN frequency for scalar($+$) perturbation in $D=5$.} 
\label{qnmS+5}
\end{table}

\begin{figure}[h!]
\noindent\begin{subfigure}[b]{0.5\textwidth}
    \centering
    \includegraphics[scale=0.52]{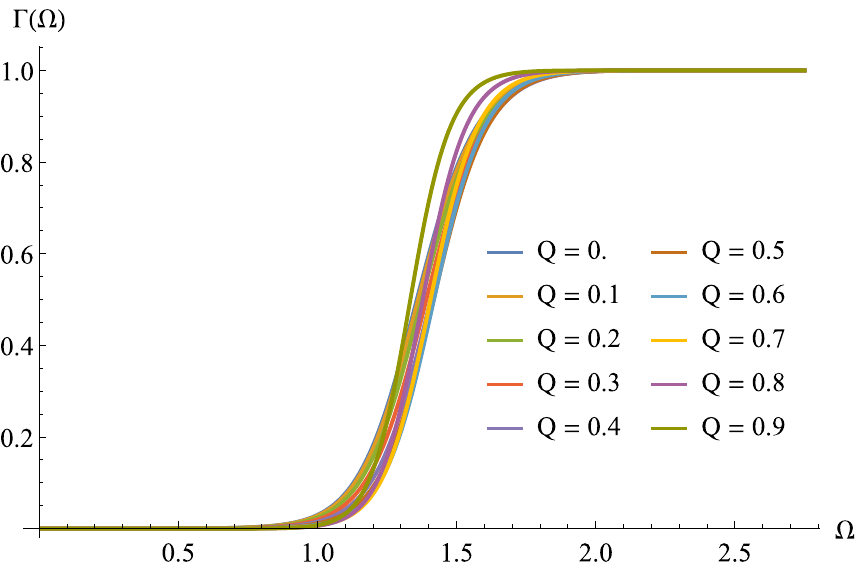}
\end{subfigure}%
\noindent\begin{subfigure}[b]{0.5\textwidth}
    \centering
    \includegraphics[scale=0.39]{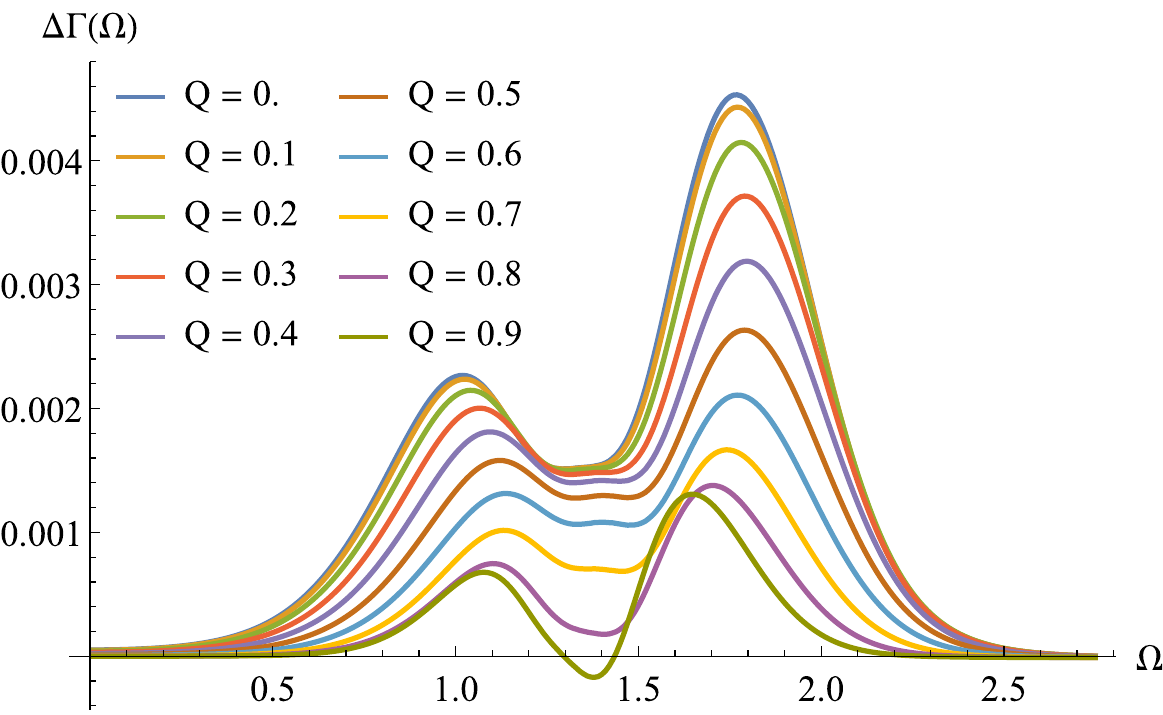}
\end{subfigure}
\caption{GBFs obtained using correspondence for scalar($+$) perturbation in $D=5$ (Left) and their differences from the numerically computed GBFs (Right).} \label{gbfS+5}
\end{figure}

Upon substituting the values of $\omega_0$ and $\omega_1$ into the formula \eqref{corr}, we obtained the GBFs for a given value of $\Omega$. The left panel of \autoref{gbfS+5} shows the GBFs determined by the correspondence for black holes with each charge $Q$. Furthermore, we numerically calculated the GBFs to compare them with those obtained from the analytical formula. The differences between the two sets of GBFs are plotted in the right panel of \autoref{gbfS+5}. For all $Q$, the differences $\Delta\Gamma(\Omega)$ remain small, implying that the correspondence achieves good accuracy for the scalar($+$) perturbation in $D=5$. As $\Delta\Gamma(\Omega)$ tends to decrease with increasing $Q$, the correspondence is more accurate for black holes with higher charge.

\begin{figure}[h!] 
\noindent\begin{subfigure}[b]{0.5\textwidth}
    \centering
    \includegraphics[scale=0.5]{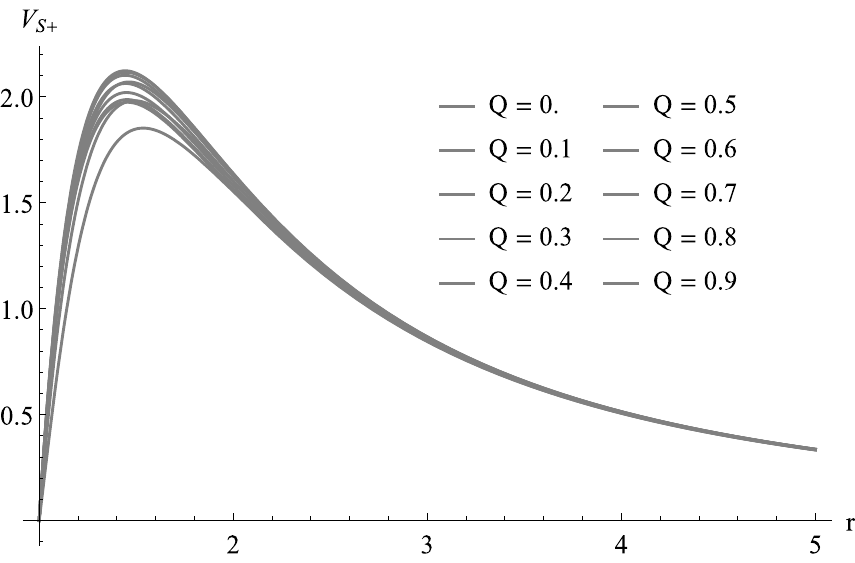}
    \caption{$D=5$}
\end{subfigure}%
\noindent\begin{subfigure}[b]{0.5\textwidth}
    \centering
    \includegraphics[scale=0.5]{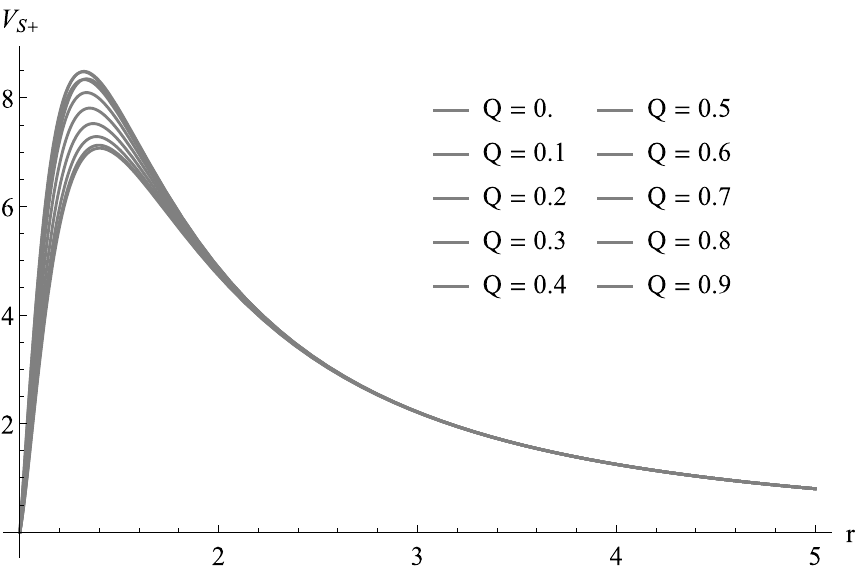}
    \caption{$D=8$}
\end{subfigure}%
\caption{Effective potentials of scalar($+$) perturbation in $D=5$ and $D=8$.} \label{VS+D}
\end{figure}

In \cite{Han:2026fpn}, the performance of the correspondence was connected with the effective potential for Schwarzschild--Tangherlini black holes with $Q=0$. We plotted the potentials $V_{S+}$ \eqref{VS} for charged black holes in \autoref{VS+D}. The graphs are shown in grey if the potential has a single maximum. We see that the potentials of scalar($+$) perturbations in $D=5$ have a single positive barrier, a suitable form for the WKB approach, leading to the remarkable performance of the correspondence. As the potentials in $D=8$ also exhibit a single peak for all $Q$, the correspondence is expected to provide a good approximation for GBFs.

\begin{figure}[h!]
\noindent\begin{subfigure}[b]{0.5\textwidth}
    \centering
    \includegraphics[scale=0.57]{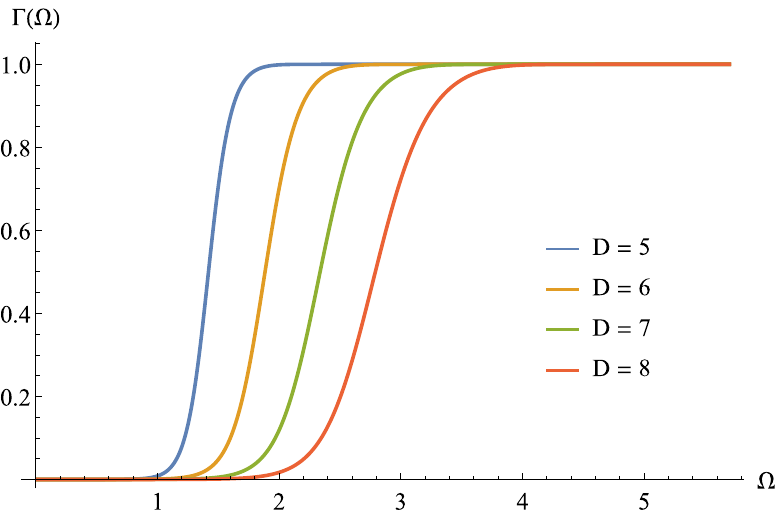}
\end{subfigure}%
\noindent\begin{subfigure}[b]{0.5\textwidth}
    \centering
    \includegraphics[scale=0.57]{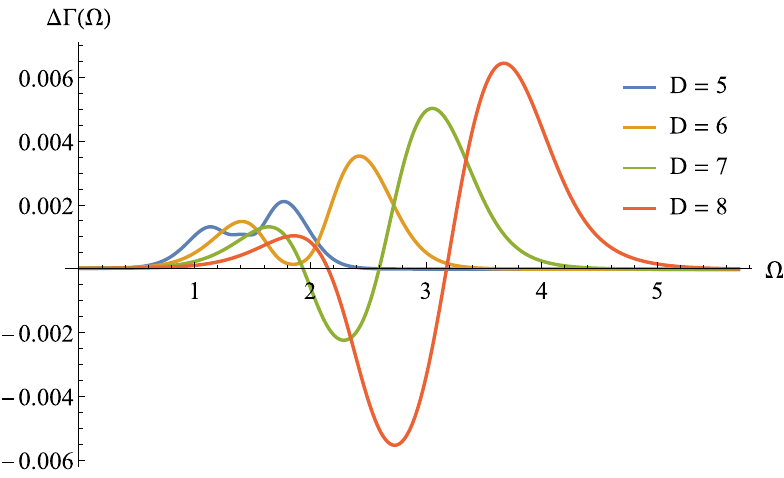}
\end{subfigure}
\caption{GBFs obtained using correspondence for scalar($+$) perturbation for $Q=0.6$ (Left) and their differences from the numerically computed GBFs (Right).} \label{gbfS+Q06}
\end{figure}

To explicitly examine its validity in higher dimensions, we calculated the GBFs for a fixed value of charge in each dimension. The QNMs computed for black holes with $Q=0.6$ are presented in \autoref{qnmS+Q06}. The approximate formula \eqref{corr} yielded the GBFs from the values of $\omega_0$ and $\omega_1$. We plotted the GBFs and their differences from those computed numerically in \autoref{gbfS+Q06}.

\begin{table}[h!]
\centering
\begin{tabular}{
|>{\centering\arraybackslash}p{1.5cm}|
 >{\centering\arraybackslash}p{4.5cm}|
 >{\centering\arraybackslash}p{4.5cm}| }

\hline\hline
\multicolumn{3}{|c|}{\textbf{$Q=0.6, \, l=2$}} \\
\hline\hline

$D$ & $n=0$ & $n=1$ \\
\hline
$5$   & $1.409221 - 0.287425 i
$ & $1.333879 - 0.878759 i$ \\
$6$   & $1.862039 - 0.423021 i
$ & $1.689232 - 1.300624 i
$ \\
$7$   & $2.297925 - 0.552149 i
$ & $1.990640 - 1.702993 i
$ \\
$8$   & $2.732166 - 0.674569 i
$ & $2.263113 - 2.077894 i
$ \\
\hline\hline

\end{tabular}
    \caption{QN frequency for scalar($+$) perturbation for $Q=0.6$.} 
\label{qnmS+Q06}
\end{table}

\begin{figure}[h!]
\noindent\begin{subfigure}[b]{0.5\textwidth}
    \centering
    \includegraphics[scale=0.52]{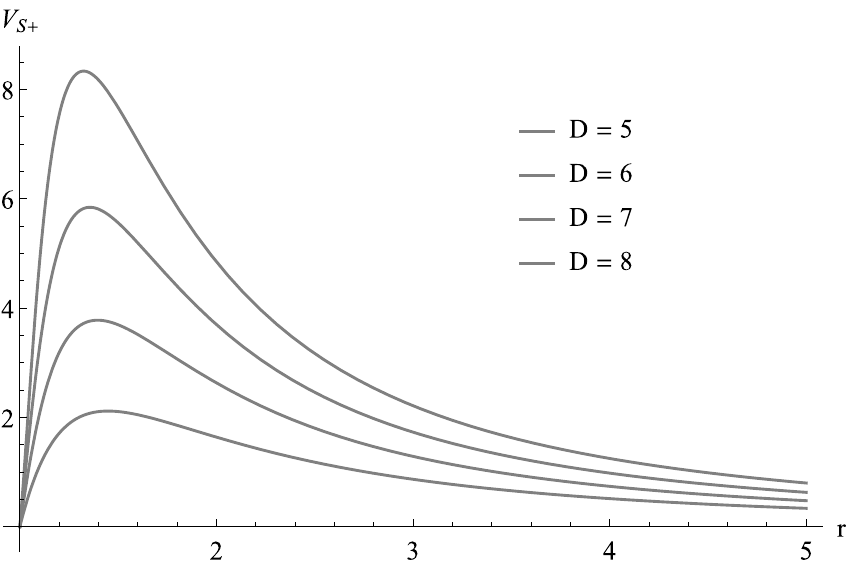}
    \caption{$Q=0.6$}
\end{subfigure}%
\noindent\begin{subfigure}[b]{0.5\textwidth}
    \centering
    \includegraphics[scale=0.52]{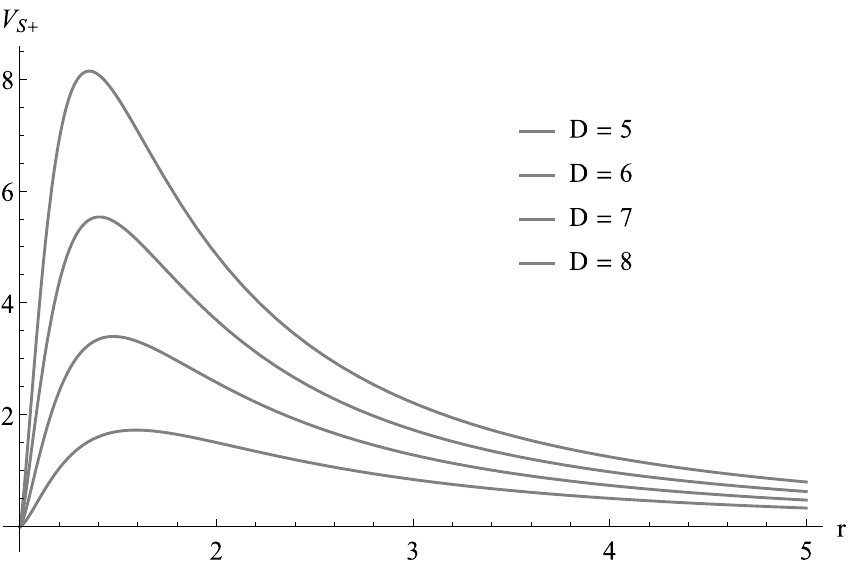}
    \caption{$Q=0.98$}
\end{subfigure}
\caption{Effective potentials of scalar($+$) perturbation for $Q=0.6$ and $Q=0.98$.} \label{VS+Q}
\end{figure}

The right panel shows that the deviations of the GBFs obtained by the correspondence from numerical GBFs are small in $D=5,6,7$, and $8$. This indicates a high precision of the correspondence for the scalar($+$) type of perturbation. As illustrated in \autoref{VS+Q}, the single-barrier structure of the potential at $Q=0.6$ in each dimension contributes to the validity of the relationship between QNMs and GBFs. The maximum magnitude of the difference increases as the dimension increases. This is likely because the higher-order WKB approach employed to establish the correspondence suffers a loss of accuracy at higher dimensions \cite{Konoplya:2003ii,Konoplya:2019hlu}. However, the maximum $|\Delta \Gamma(\Omega)|$ in $D=8$ remains below $0.007$, indicating a reliable approximation for the GBF.

\begin{figure}[h!]
\noindent\begin{subfigure}[b]{0.5\textwidth}
    \centering
    \includegraphics[scale=0.52]{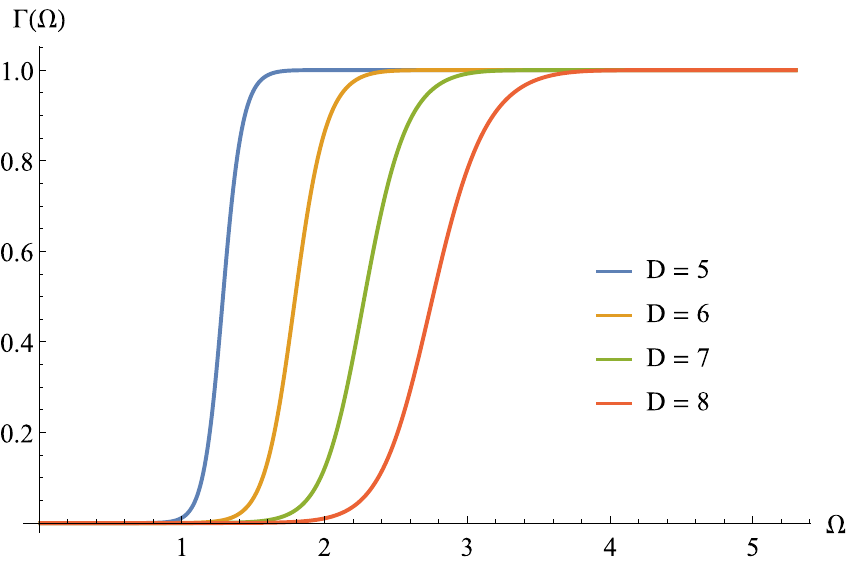}
\end{subfigure}%
\noindent\begin{subfigure}[b]{0.5\textwidth}
    \centering
    \includegraphics[scale=0.52]{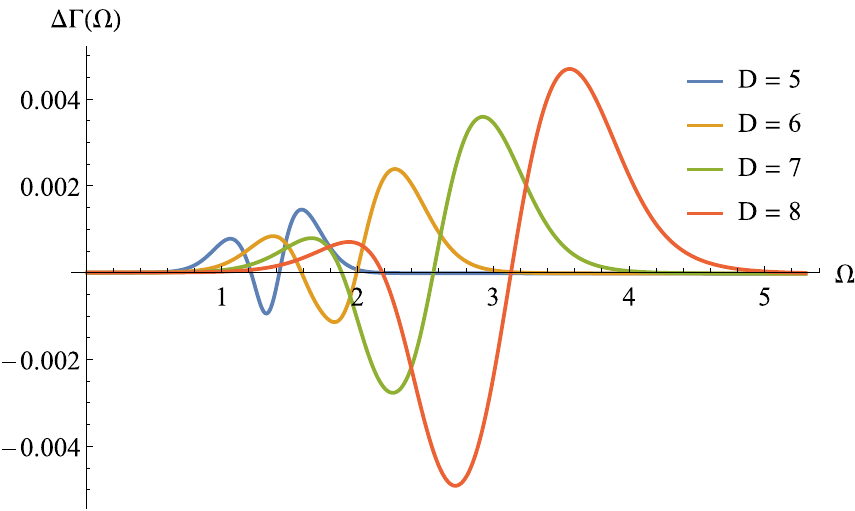}
\end{subfigure}
\caption{GBFs obtained using correspondence for scalar($+$) perturbation for $Q=0.98$ (Left) and their differences from the numerically computed GBFs (Right).} \label{gbfS+Q098}
\end{figure}

Furthermore, we obtained the GBFs for near-extreme black holes with $Q=0.98$. In the near-extreme regime, the singularities in the wave equation \eqref{waveeq} are located close to $\rho=1$, and thus the convergence of the series in Eq. \eqref{ansatz} deteriorates. Hence, we continued the series by introducing one or two midpoints and substantially increased the depth of the continued fraction to compute the QNMs of the near-extreme black holes. Our results for the $n=0$ and $n=1$ modes are given in \autoref{qnmS+Q098}. We then substituted them into Eq. \eqref{corr} to find the GBFs. The plots of the GBFs obtained by the correspondence and the discrepancy between the approximate and the numerical values are presented in \autoref{gbfS+Q098}. As indicated by the small values of $|\Delta \Gamma(\Omega)|$, the correspondence for near-extreme black holes provides a reliable approximation for GBFs in all cases. This is attributable to the effective potentials for which the WKB method works well, as shown in the right panel of \autoref{VS+Q}.

\begin{table}[h!]
\centering
\begin{tabular}{
|>{\centering\arraybackslash}p{1.5cm}|
 >{\centering\arraybackslash}p{4.5cm}|
 >{\centering\arraybackslash}p{4.5cm}| }

\hline\hline
\multicolumn{3}{|c|}{\textbf{$Q=0.98, \, l=2$}} \\
\hline\hline

$D$ & $n=0$ & $n=1$ \\
\hline
$5$   & $1.279325 - 0.214415 i
$ & $1.209499 - 0.651731 i$ \\
$6$   & $1.775009  -0.341056 i
$ & $1.618102 - 1.037907 i
$ \\
$7$   & $2.246018 - 0.463818 i
$ & $1.974841 - 1.408993 i
$ \\
$8$   & $2.708382 - 0.581263 i
$ & $2.304260 - 1.755711 i
$ \\
\hline\hline

\end{tabular}
    \caption{QN frequency for scalar($+$) perturbation for $Q=0.98$.} 
\label{qnmS+Q098}
\end{table}

\subsection{Scalar($-$) perturbation}
We present the performance of the correspondence for the scalar($-$) type of gravitational perturbation. This type reduces to the scalar gravitational perturbation of Schwarzschild--Tangherlini black holes in the limit of zero charge. The correspondence for the scalar type was previously found in \cite{Han:2026fpn} to have a behavior distinct from that for other types of perturbation. As shown below, scalar($-$) perturbations of charged black holes also exhibit more complex behavior than that of other perturbations. 
 
We numerically computed the QNMs to calculate the GBFs through the correspondence. The frequencies $\omega_0$ and $\omega_1$ of five-dimensional black holes with each value of charge $Q$ are shown in \autoref{qnmS-5}. The fundamental modes presented in this subsection are consistent with those obtained in \cite{Konoplya:2008au}. Unlike those for the scalar($+$) type, the magnitudes of the real and imaginary parts decrease monotonically with increasing $Q$.

\begin{table}[h!]
\centering
\begin{tabular}{
|>{\centering\arraybackslash}p{1.5cm}|
 >{\centering\arraybackslash}p{4.5cm}|
 >{\centering\arraybackslash}p{4.5cm}| }

\hline\hline
\multicolumn{3}{|c|}{\textbf{$D=5, \, l=2$}} \\
\hline\hline

$Q$ & $n=0$ & $n=1$ \\
\hline
$0$   & $0.947739 - 0.256094 i$ & $0.851234 - 0.821160 i$ \\
$0.1$   & $0.940946 - 0.253711 i$ & $0.846265 - 0.813877 i$ \\
$0.2$   & $0.921884 - 0.247013 i$ & $0.832156 - 0.793185 i$ \\
$0.3$   & $0.893564 - 0.237099 i$ & $0.810529 - 0.761892 i$ \\
$0.4$   & $0.859131 - 0.225284 i$ & $0.782591 - 0.723412 i$ \\
$0.5$   & $0.821129 - 0.212860 i$ & $0.748667 - 0.681530 i$ \\
$0.6$   & $0.781586 - 0.200991 i$ & $0.709002 - 0.641205 i$ \\
$0.7$   & $0.742389 - 0.190461 i$ & $0.667711 - 0.608744 i$ \\
$0.8$   & $0.705137 - 0.181074 i$ & $0.634010 - 0.580866 i$ \\
$0.9$   & $0.670130 - 0.172135 i$ & $0.602663 - 0.552021 i$ \\
\hline\hline

\end{tabular}
    \caption{QN frequency for scalar($-$) perturbation in $D=5$.} 
\label{qnmS-5}
\end{table}

\begin{figure}[h!]
\noindent\begin{subfigure}[b]{0.5\textwidth}
    \centering
    \includegraphics[scale=0.5]{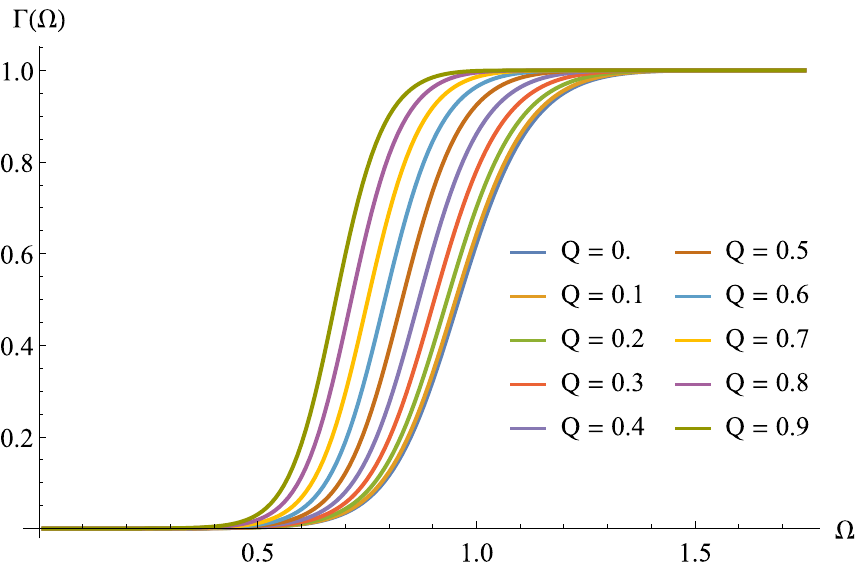}
\end{subfigure}%
\noindent\begin{subfigure}[b]{0.5\textwidth}
    \centering
    \includegraphics[scale=0.41]{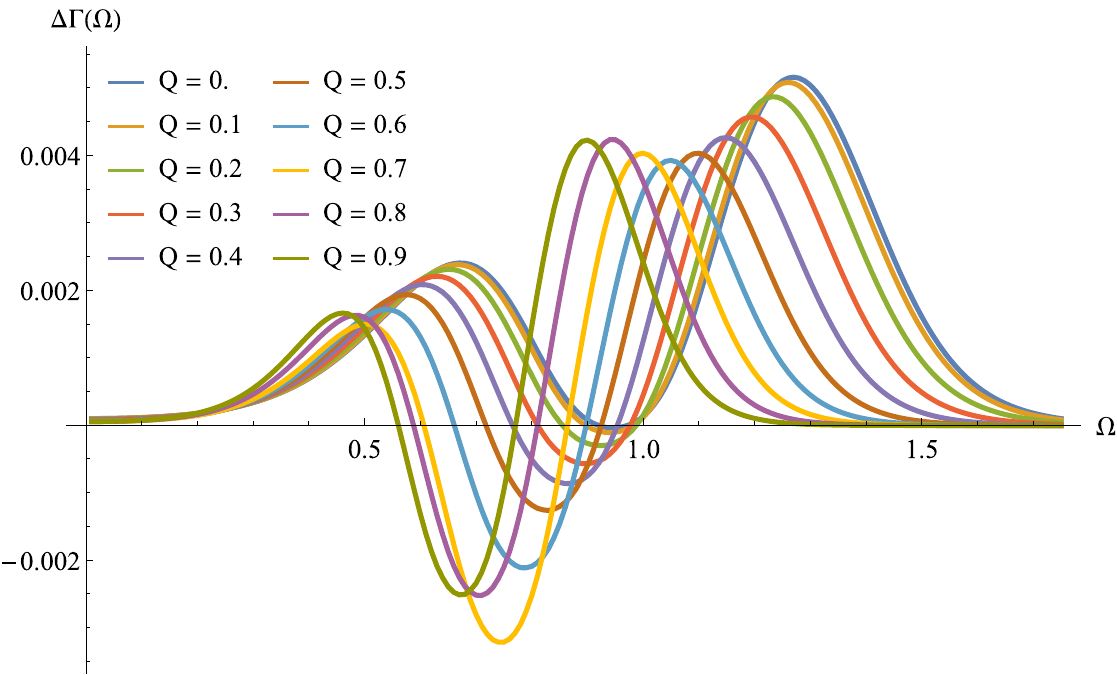}
\end{subfigure}
\caption{GBFs obtained using correspondence for scalar($-$) perturbation in $D=5$ (Left) and their differences from the numerically computed GBFs (Right).} \label{gbfS-5}
\end{figure}

By substituting the values into Eq. \eqref{corr}, we obtained the GBF at a given frequency $\Omega$. The left panel of \autoref{gbfS-5} shows the approximate GBFs calculated using the correspondence. We also computed the GBFs using a numerical method and plotted their differences from the approximate values in the right panel. The small differences $\Delta\Gamma(\Omega)$ imply that the correspondence for the scalar($-$) perturbation achieves high precision for all values of $Q$ in $D=5$.

\begin{figure}[h!] 
\noindent\begin{subfigure}[b]{0.5\textwidth}
    \centering
    \includegraphics[scale=0.5]{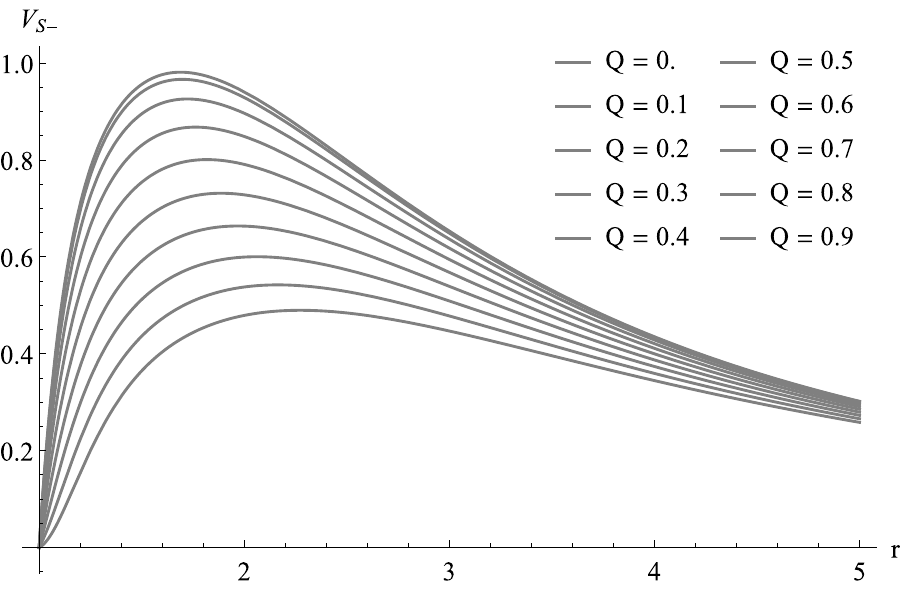}
    \caption{$D=5$} \label{fig7a}
\end{subfigure}%
\noindent\begin{subfigure}[b]{0.5\textwidth}
    \centering
    \includegraphics[scale=0.5]{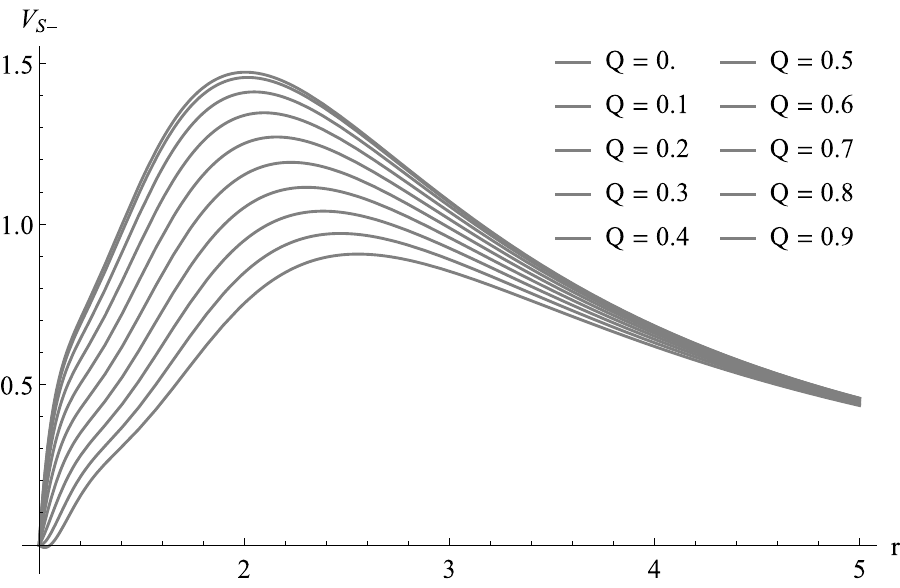}
    \caption{$D=6$}
\end{subfigure} \\

\noindent\begin{subfigure}[b]{0.5\textwidth}
    \centering
    \includegraphics[scale=0.5]{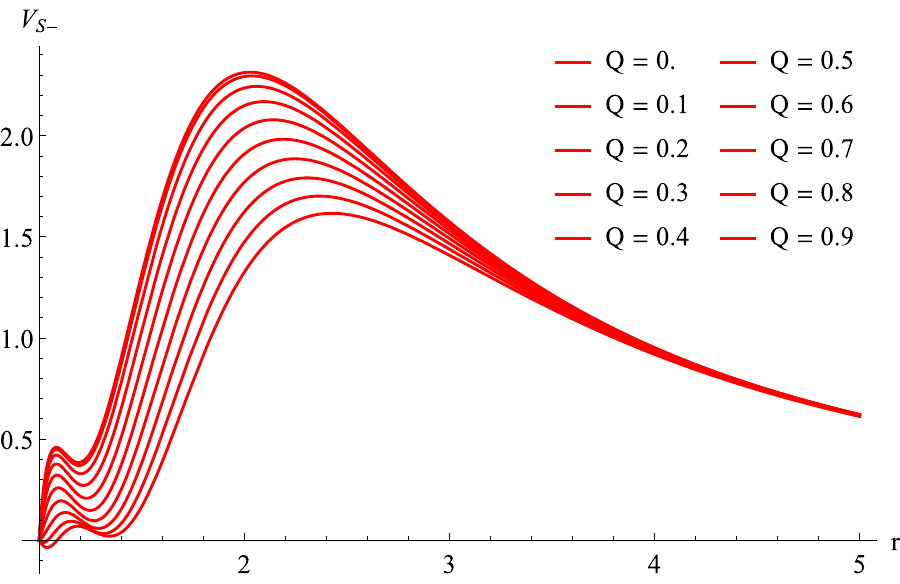}
    \caption{$D=7$}
\end{subfigure}%
\noindent\begin{subfigure}[b]{0.5\textwidth}
    \centering
    \includegraphics[scale=0.5]{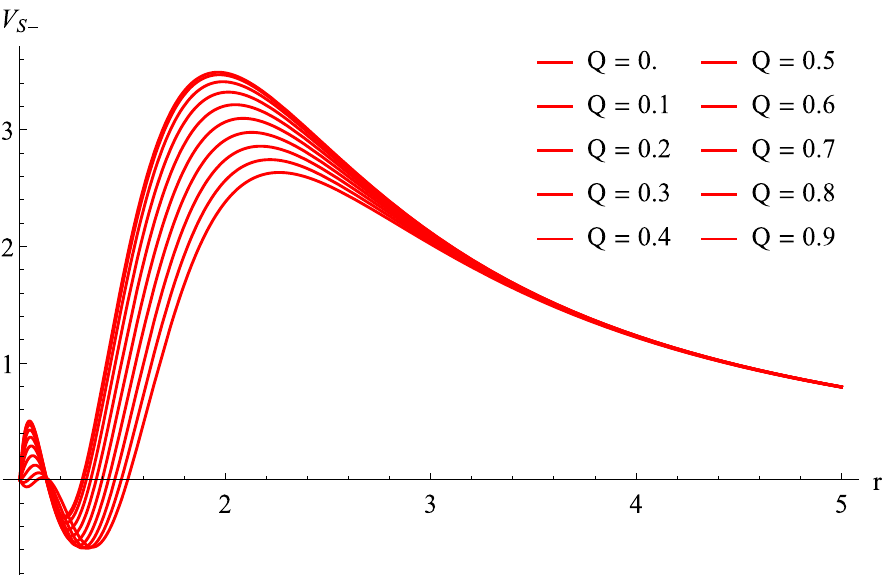}
    \caption{$D=8$}
\end{subfigure}%
\caption{Effective potentials of scalar($-$) perturbation in $D=5,6,7$, and $8$.} \label{VSplotD}
\end{figure}

We illustrated the effective potentials $V_{S-}$ of scalar($-$) perturbations, given by Eq. \eqref{VS}, for $D=5,6,7$, and $8$ in \autoref{VSplotD}. Potentials with a single maximum are plotted in grey, while those with two or more maxima are plotted in red. \autoref{fig7a} shows that the five-dimensional black holes with all values of the charge have a single-peak potential. Such a potential belongs to the class for which the WKB approach works well, leading to the good performance of the correspondence as shown in \autoref{gbfS-5}. The potentials also exhibit a single barrier for all $Q$ in $D=6$ but develop a second barrier near the event horizon in $D=7$ and $D=8$. We examined the correspondence for $D=7$, where multiple peaks first appear.

We employed the integration-through-midpoints method to compute the QNMs of seven-dimensional black holes. A midpoint $\rho_0$ was used to construct the series as in Eq. \eqref{news} for black holes with charge $Q<0.8$, whereas two midpoints, $\rho_0$ and $\rho_1$, were needed for higher charges for the series to converge between the event horizon and infinity. The results are presented in \autoref{qnmS-7}. As in the case of $D=5$, the real parts decrease monotonically with growing $Q$, except for the high-$Q$ regime of the first overtone. However, the imaginary parts of both $\omega_0$ and $\omega_1$ do not exhibit monotonic behavior. 

\begin{table}[h!]
\centering
\begin{tabular}{
|>{\centering\arraybackslash}p{1.5cm}|
 >{\centering\arraybackslash}p{4.5cm}|
 >{\centering\arraybackslash}p{4.5cm}| }

\hline\hline
\multicolumn{3}{|c|}{\textbf{$D=7, \, l=2$}} \\
\hline\hline

$Q$ & $n=0$ & $n=1$ \\
\hline
$0$   & $1.339158 - 0.400860 i$ & $1.185695 - 1.100378 i$ \\
$0.1$   & $1.331837 - 0.400562 i$ & $1.179698 - 1.090348 i$ \\
$0.2$   & $1.311363 - 0.400306 i$ & $1.161451 - 1.061660 i$ \\
$0.3$   & $1.281662 - 0.401696 i$ & $1.129749 - 1.018067 i$ \\
$0.4$   & $1.248303 - 0.406452 i$ & $1.081359 - 0.965418 i$ \\
$0.5$   & $1.218338 - 0.414393 i$ & $1.010141 - 0.914494 i$ \\
$0.6$   & $1.197456 - 0.419240 i$ & $0.912490 - 0.891908 i$ \\
$0.7$   & $1.180123 - 0.412358 i$ & $0.842223 - 0.947155 i$ \\
$0.8$   & $1.156032 - 0.398594 i$ & $0.865678 - 0.946362 i$ \\
$0.9$   & $1.127843 - 0.387023 i$ & $0.845565 - 0.919864 i$ \\
\hline\hline

\end{tabular}
    \caption{QN frequency for scalar($-$) perturbation in $D=7$.} 
\label{qnmS-7}
\end{table}

\begin{figure}[h!]
\noindent\begin{subfigure}[b]{0.5\textwidth}
    \centering
    \includegraphics[scale=0.48]{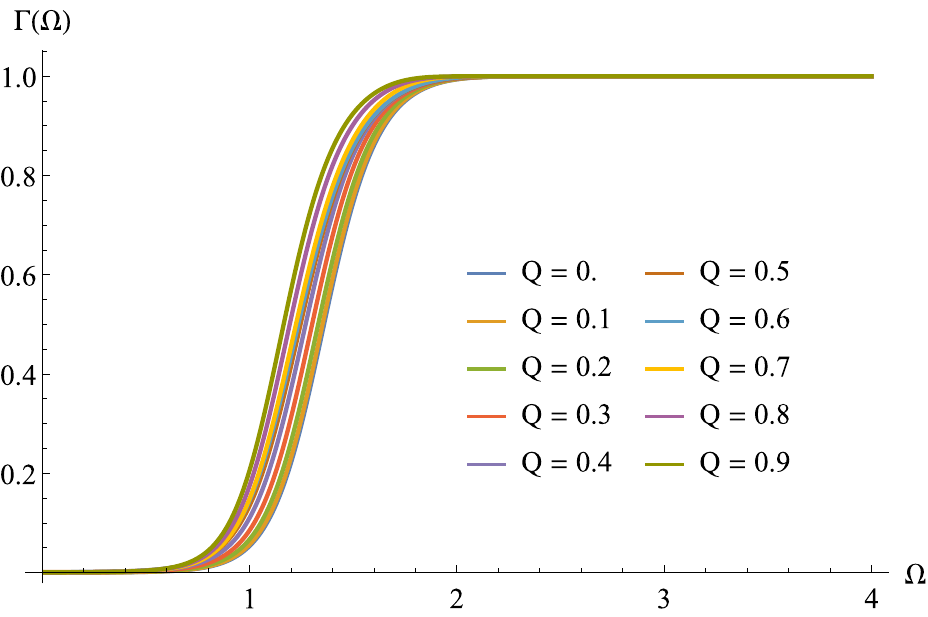}
\end{subfigure}%
\noindent\begin{subfigure}[b]{0.5\textwidth}
    \centering
    \includegraphics[scale=0.48]{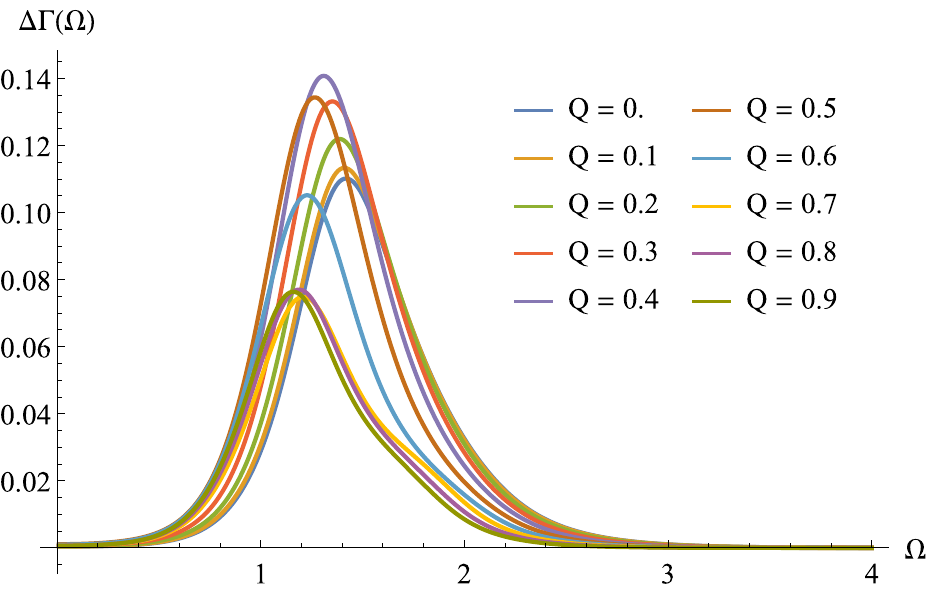}
\end{subfigure}
\caption{GBFs obtained using correspondence for scalar($-$) perturbation in $D=7$ (Left) and their differences from the numerically computed GBFs (Right).} \label{gbfS-7}
\end{figure}

Using the QN frequencies, we obtained the GBFs through the correspondence equation \eqref{corr}. \autoref{gbfS-7} shows the approximate GBFs and the discrepancy between them and the accurate GBFs computed numerically. The maximum of the difference $|\Delta \Gamma(\Omega)|$ grows from the $Q=0$ case, which corresponds to the Schwarzschild--Tangherlini black hole, to the $Q=0.4$ case. These cases exhibit a maximum deviation of more than ten percent, considerably larger than that in $D=5$, indicating that the performance of the correspondence deteriorates in this regime. Then, for $Q > 0.4$, the maximum $|\Delta \Gamma(\Omega)|$ decreases as $Q$ increases and remains below $0.08$ for $Q > 0.6$. Although the effective potential possesses the double-peak form that lies beyond the assumption of the WKB approach, the correspondence recovers reasonable accuracy in some cases.

To further examine this behavior, we calculated the GBFs for a fixed $Q$ in higher dimensions. For example, we found the QNMs of black holes with charge $Q=0.6$ in $D=5,6,7$, and $8$ to determine the GBFs from the correspondence. The integration-through-midpoints method was applied to compute the QN frequencies for $D=7$ and $D=8$, and the QN frequencies for all four dimensions are given in \autoref{qnmS-Q06}.

\begin{table}[h!]
\centering
\begin{tabular}{
|>{\centering\arraybackslash}p{1.5cm}|
 >{\centering\arraybackslash}p{4.5cm}|
 >{\centering\arraybackslash}p{4.5cm}| }

\hline\hline
\multicolumn{3}{|c|}{\textbf{$Q=0.6, \, l=2$}} \\
\hline\hline

$D$ & $n=0$ & $n=1$ \\
\hline
$5$   & $0.781586 - 0.200991 i
$ & $0.709002 - 0.641205 i$ \\
$6$   & $0.970219 - 0.271423 i
$ & $0.839081 - 0.826627 i
$ \\
$7$   & $1.197456 - 0.419240 i
$ & $0.912490 - 0.891908 i
$ \\
$8$   & $1.545048 - 0.623475 i
$ & $0.882550 - 0.842976 i
$ \\
\hline\hline

\end{tabular}
    \caption{QN frequency for scalar($-$) perturbation for $Q=0.6$.} 
\label{qnmS-Q06}
\end{table}

\begin{figure}[h!]
\noindent\begin{subfigure}[b]{0.5\textwidth}
    \centering
    \includegraphics[scale=0.51]{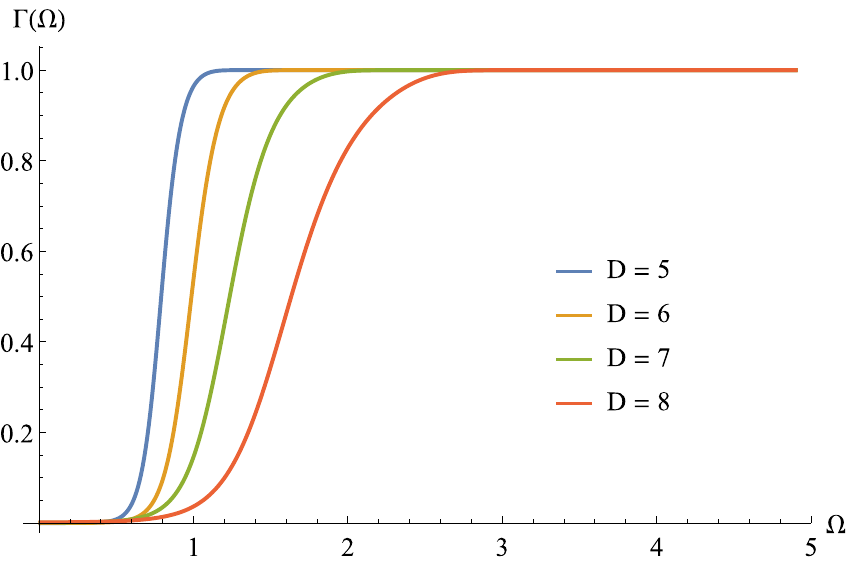}
\end{subfigure}%
\noindent\begin{subfigure}[b]{0.5\textwidth}
    \centering
    \includegraphics[scale=0.51]{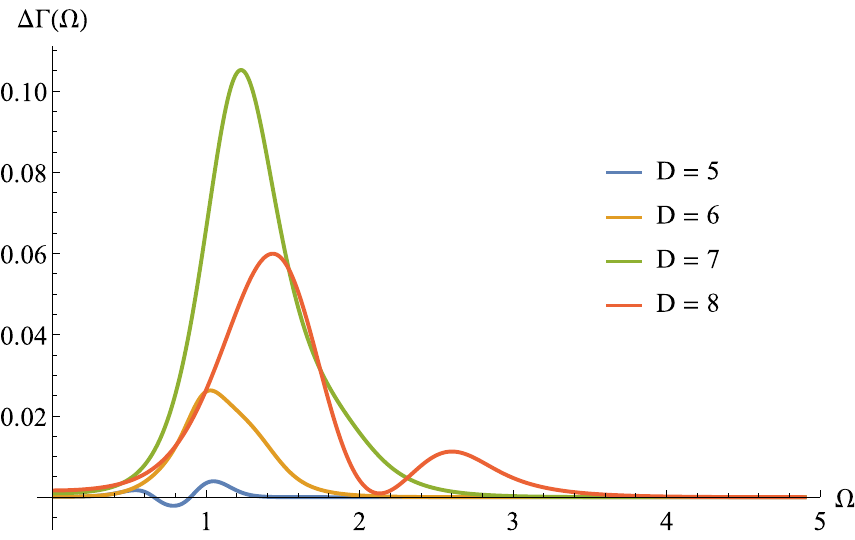}
\end{subfigure}
\caption{GBFs obtained using correspondence for scalar($-$) perturbation for $Q=0.6$ (Left) and their differences from the numerically computed GBFs (Right).
} \label{gbfS-Q06}
\end{figure}

We substituted $\omega_0$ and $\omega_1$ into Eq. \eqref{corr} and plotted the resulting GBFs in \autoref{gbfS-Q06}. We also presented the discrepancy between the GBFs obtained by the correspondence and those obtained by numerical computation. For $D=5$ and $D=6$, indicated by the blue and yellow curves, respectively, the values of $\Delta\Gamma(\Omega)$ are relatively small, suggesting that the correspondence offers a reliable approximation for GBFs. 

For $D=7$, the green curve shows that the maximum magnitude of the discrepancy is markedly larger than that for $D=5$ and $D=6$. The approximate values deviate considerably from the accurate values, indicating that the relation between QNMs and GBFs does not hold for the seven-dimensional black hole with $Q=0.6$.

\begin{figure}[h!]
\noindent\begin{subfigure}[b]{0.5\textwidth}
    \centering
    \includegraphics[scale=0.52]{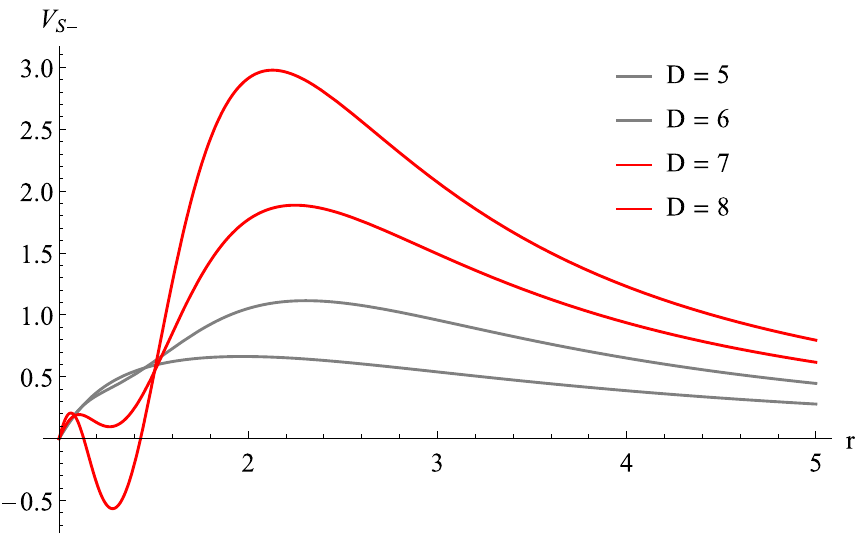}
    \caption{$Q=0.6$}
\end{subfigure}%
\noindent\begin{subfigure}[b]{0.5\textwidth}
    \centering
    \includegraphics[scale=0.52]{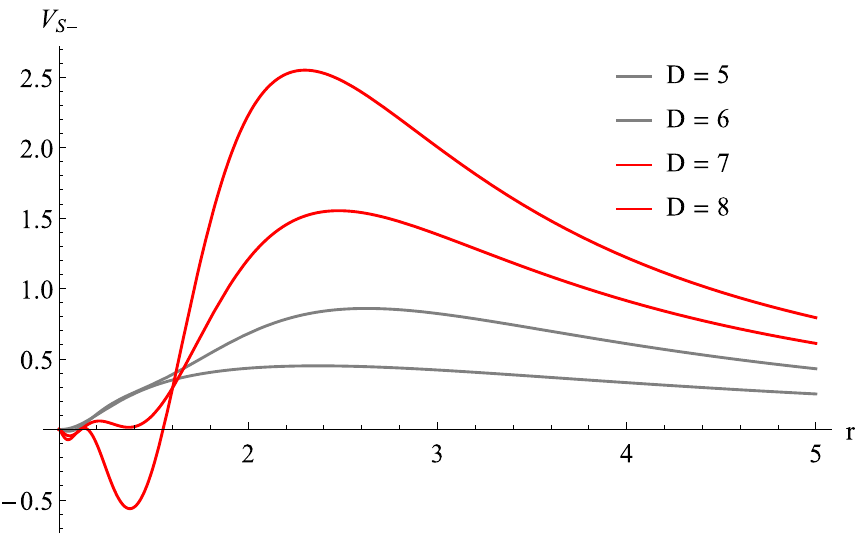}
    \caption{$Q=0.98$}
\end{subfigure} 
\caption{Effective potentials of scalar($-$) perturbation for $Q=0.6$ and $Q=0.98$.} \label{VS-Q}
\end{figure}

However, the difference decreases for $D=8$, as represented by the red curve. The accuracy of the GBFs obtained by the correspondence is considerably improved relative to that for $D=7$. This phenomenon is unique to the scalar($-$) gravitational perturbation. As shown in \autoref{gbfS+Q06}, the correspondence for the scalar($+$) type loses accuracy as the dimension increases. Since the changes across dimensions were continuous, we speculated that this trend is due to an inherent property of the higher-order WKB approximation, whose errors grow at higher dimensions. However, the scalar($-$) type of perturbation does not follow this trend. We plotted the effective potential with $Q=0.6$ for each dimension in the left panel of \autoref{VS-Q}. Potentials with multiple peaks are represented by the red lines. Although the potentials for both $D=7$ and $D=8$ clearly have two peaks, the accuracy of the correspondence increased from $D=7$ to $D=8$.

Furthermore, the magnitude of the difference $|\Delta \Gamma (\Omega)|$ in $D=8$ is not sufficiently large to conclude that the correspondence is invalid. As in the high-$Q$ regime in $D=7$, the GBFs obtained from the correspondence achieve moderate accuracy, even for the potential that deviates from the form assumed in the standard WKB approach. Therefore, we refrain from making a categorical assessment of the correspondence in these ambiguous cases and report only quantitative results of its accuracy.

\begin{figure}[h!]
\noindent\begin{subfigure}[b]{0.5\textwidth}
    \centering
    \includegraphics[scale=0.52]{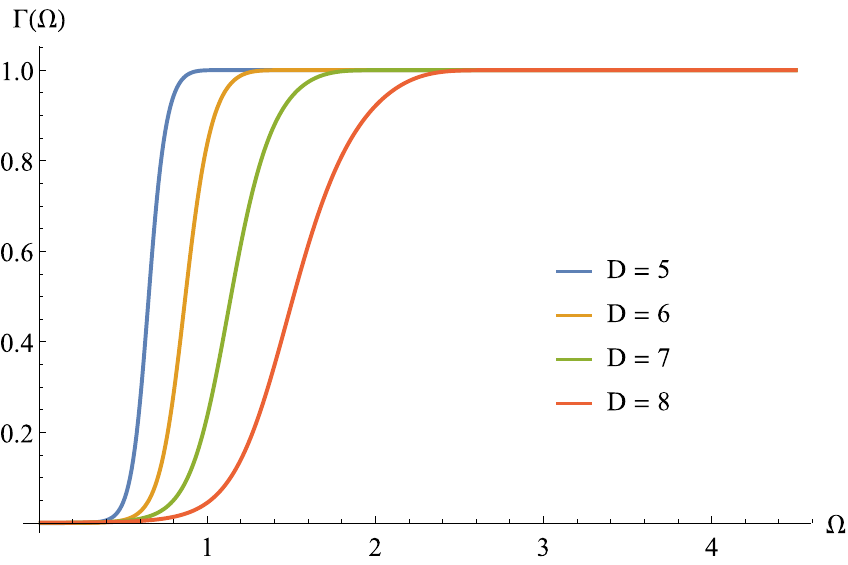}
\end{subfigure}%
\noindent\begin{subfigure}[b]{0.5\textwidth}
    \centering
    \includegraphics[scale=0.52]{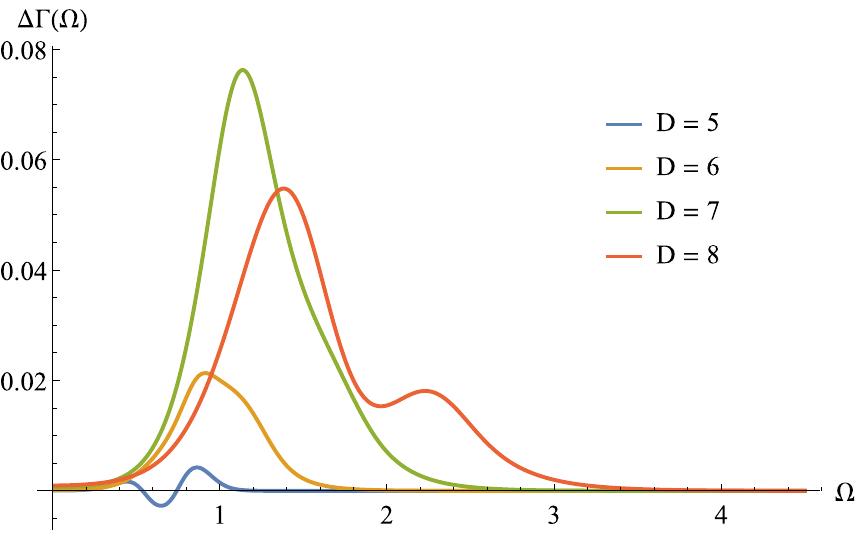}
\end{subfigure}
\caption{GBFs obtained using correspondence for scalar($-$) perturbation for $Q=0.98$ (Left) and their differences from the numerically computed GBFs (Right).
} \label{gbfS-Q098}
\end{figure}

We also investigated the correspondence for near-extreme cases. To compute the QN frequencies of black holes with $Q=0.98$, we introduced two midpoints $\rho_0$ and $\rho_1$ to expand the series as in Eq. \eqref{news} and increased the depth of the continued fraction. Our results for $\omega_0$ and $\omega_1$ are presented in \autoref{qnmS-Q098}. We then calculated the approximate GBFs from the correspondence \eqref{corr} and the reference GBFs using the numerical method. \autoref{gbfS-Q098} illustrates the approximate GBFs and their differences from the numerically computed GBFs.

\begin{table}[h!]
\centering
\begin{tabular}{
|>{\centering\arraybackslash}p{1.5cm}|
 >{\centering\arraybackslash}p{4.5cm}|
 >{\centering\arraybackslash}p{4.5cm}| }

\hline\hline
\multicolumn{3}{|c|}{\textbf{$Q=0.98, \, l=2$}} \\
\hline\hline

$D$ & $n=0$ & $n=1$ \\
\hline
$5$   & $0.643626 - 0.165328 i
$ & $0.578828 - 0.530191 i$ \\
$6$   & $0.854710 - 0.244925 i
$ & $0.733131 - 0.757077 i
$ \\
$7$   & $1.105518 - 0.379003 i
$ & $0.829412 - 0.901894 i
$ \\
$8$   & $1.435632 - 0.552161 i
$ & $0.841175 - 0.969673 i
$ \\
\hline\hline

\end{tabular}
    \caption{QN frequency for scalar($-$) perturbation for $Q=0.98$.} 
\label{qnmS-Q098}
\end{table}

The small differences $\Delta \Gamma(\Omega)$ for $D=5$ and $D=6$ indicate good performance of the correspondence, as observed for $Q=0.6$. For $D=7$ and $D=8$, the GBFs obtained by the correspondence exhibit reasonable accuracy even though the potential has double maxima, as shown in the right panel of \autoref{VS-Q}. For near-extreme black holes, the accuracy also tends to improve from $D=7$ to $D=8$. These results were obtained only for the scalar($-$) type of perturbation. 

Moreover, as shown in \autoref{qnmS-Q06}, we observed that the absolute values of the real and imaginary parts of $\omega_1$ increase with $D$ for fixed $Q=0.6$. However, both quantities decrease at $D=8$. We further calculated the QNMs for $D=9$ and $D=10$, obtaining $ \omega_0=1.959555 - 0.763783 i$ and $ \omega_1=0.816447 - 0.823960 i$ for $D=9$, and $ \omega_0=2.379521 - 0.875468 i$ and $ \omega_1=0.770497 - 0.813677 i$ for $D=10$.
The magnitudes of the real and imaginary parts of $\omega_1$ continue to decrease up to $D=10$, where the magnitude of its imaginary part falls below that of $\omega_0$.
We then calculated the GBFs using the QNMs for $D=9$ and $10$ to evaluate the precision of the correspondence. The maximum deviation $|\Delta \Gamma(\Omega)|$ from the accurate values further decreases at $D=9$ but increases at $D=10$. Its dependence on the dimension $D$ is therefore not monotonic, unlike the trend for the other perturbation types.
This non-monotonic dependence suggests that the unusual QNM behavior of scalar($-$) gravitational perturbations may be associated with the distinctive behavior of the GBFs obtained through the correspondence. Nevertheless, the origin of this behavior remains unclear, and we therefore refrain from further interpretation and treat it as an observed characteristic.

\section{Correspondence in vector-type perturbations}
Next, we analyze the correspondence between QNMs and GBFs for vector types of perturbation, governed by the wave equation with the effective potential \eqref{VV}. The QNMs are computed using the continued fraction and the integration-through-midpoints methods. Our results for the fundamental QNMs of the vector perturbations agree with those reported in \cite{Konoplya:2007jv}. We obtain the GBFs for non-extreme black holes with $0\le Q<1$ through the correspondence at the low multipole $l=2$ and evaluate the accuracy of these GBFs by comparing them with numerically computed values. 

\subsection{Vector($+$) perturbation}
To find the GBFs from the correspondence for the vector($+$) type of perturbation, we solved Eq. \eqref{waveeq} with the potential $V_{V+}$ and obtained the fundamental and the first overtone modes. The results for five-dimensional black holes are given in \autoref{qnmV+5}. As in the scalar($+$) perturbation, the real components of both $\omega_0$ and $\omega_1$ grow and then diminish as $Q$ increases. The decay rates decrease monotonically with increasing $Q$. 

\begin{table}[h!]
\centering
\begin{tabular}{
|>{\centering\arraybackslash}p{1.5cm}|
 >{\centering\arraybackslash}p{4.5cm}|
 >{\centering\arraybackslash}p{4.5cm}| }

\hline\hline
\multicolumn{3}{|c|}{\textbf{$D=5, \, l=2$}} \\
\hline\hline

$Q$ & $n=0$ & $n=1$ \\
\hline
$0$   & $1.468541 - 0.352426 i
$ & $1.348493 - 1.089553 i$ \\
$0.1$   & $1.471093 - 0.350744 i
$ & $1.353482 - 1.083954 i$ \\
$0.2$   & $1.477711 - 0.345599 i
$ & $1.367017 - 1.066868 i$ \\
$0.3$   & $1.485745 - 0.336755 i
$ & $1.385400 - 1.037652 i$ \\
$0.4$   & $1.491808 - 0.324002 i
$ & $1.403763 - 0.995793 i $  \\
$0.5$   & $1.492319 - 0.307338 i
$ & $1.416678 - 0.941492 i$ \\
$0.6$   & $1.483703 - 0.287232 i
$ & $1.418274 - 0.876461 i
$ \\
$0.7$   & $1.462629 - 0.264990 i
$ & $1.402632 - 0.805323 i$ \\
$0.8$   & $1.426787 - 0.243090 i
$ & $1.366358 - 0.737106 i
$ \\
$0.9$   & $1.376373 - 0.224663 i
$ & $1.313508 - 0.681916 i$ \\
\hline\hline

\end{tabular}
    \caption{QN frequency for vector($+$) perturbation in $D=5$.} 
\label{qnmV+5}
\end{table}

\begin{figure}[h!]
\noindent\begin{subfigure}[b]{0.5\textwidth}
    \centering
    \includegraphics[scale=0.45]{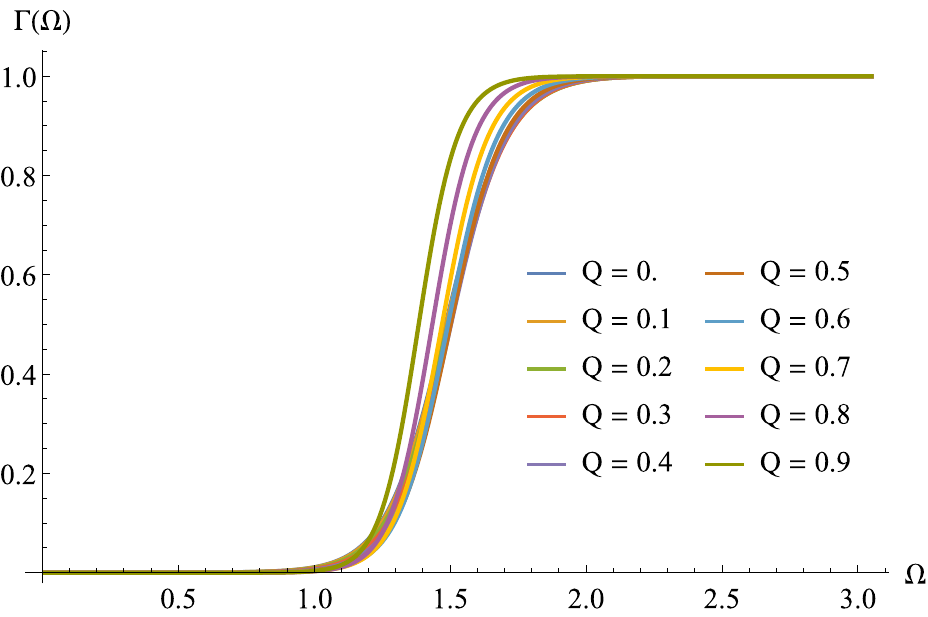}
\end{subfigure}%
\noindent\begin{subfigure}[b]{0.5\textwidth}
    \centering
    \includegraphics[scale=0.45]{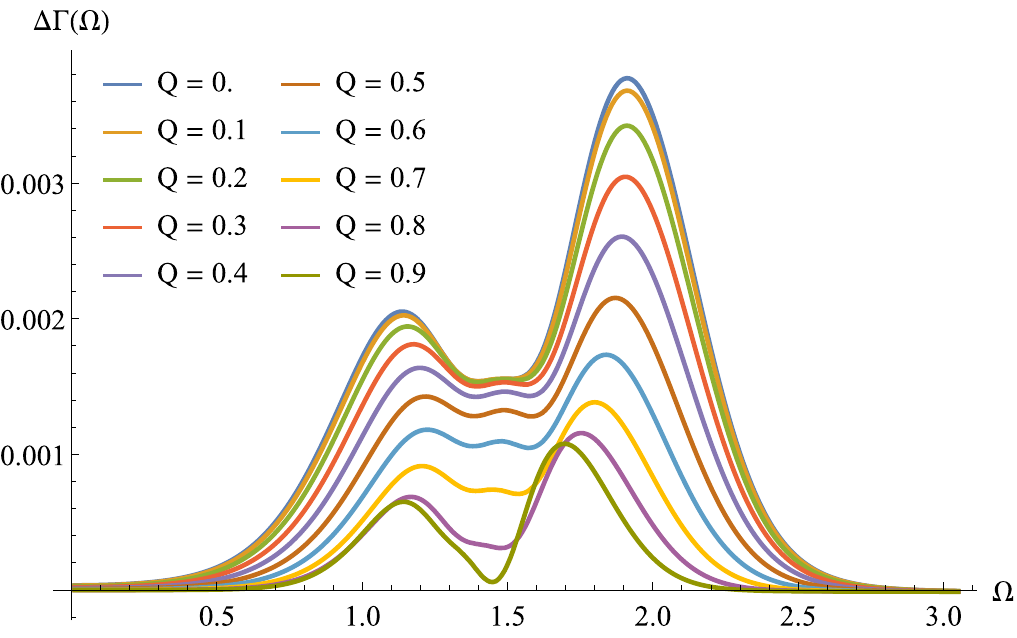}
\end{subfigure}
\caption{GBFs obtained using correspondence for vector($+$) perturbation in $D=5$ (Left) and their differences from the numerically computed GBFs (Right).} \label{gbfV+5}
\end{figure}

The GBFs can be analytically obtained by substituting $\omega_0$ and $\omega_1$ into the relation \eqref{corr} for a given value of $\Omega$. We plotted the GBFs $\Gamma(\Omega)$ for $D=5$ in \autoref{gbfV+5}. We also numerically computed the GBFs as reference values and calculated the differences $\Delta \Gamma(\Omega)$ between the analytical and numerical values, quantifying the accuracy of the correspondence. The right panel illustrates minor discrepancies for all charges, implying high accuracy of the analytical GBFs in $D=5$. The correspondence tends to be more accurate as the charge of black holes increases.

\begin{figure}[h!] 
\noindent\begin{subfigure}[b]{0.5\textwidth}
    \centering
    \includegraphics[scale=0.5]{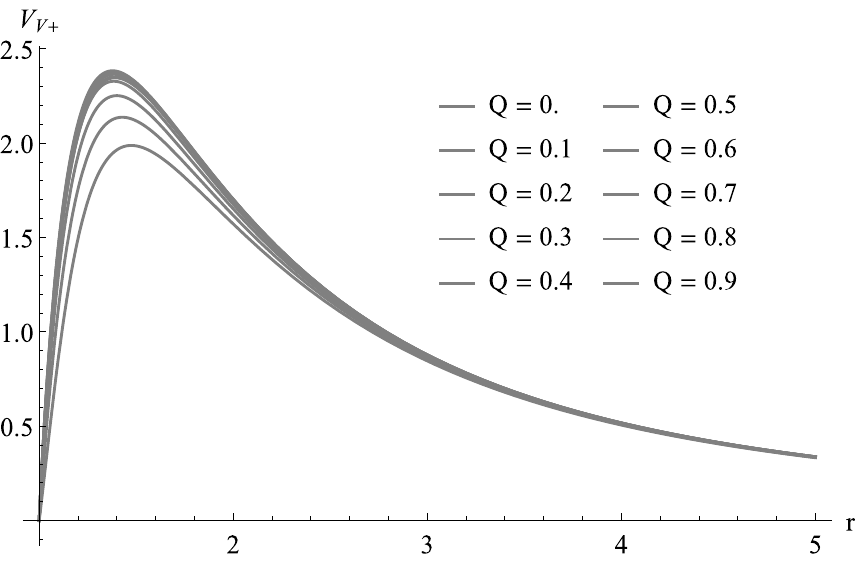}
    \caption{$D=5$}
\end{subfigure}%
\noindent\begin{subfigure}[b]{0.5\textwidth}
    \centering
    \includegraphics[scale=0.5]{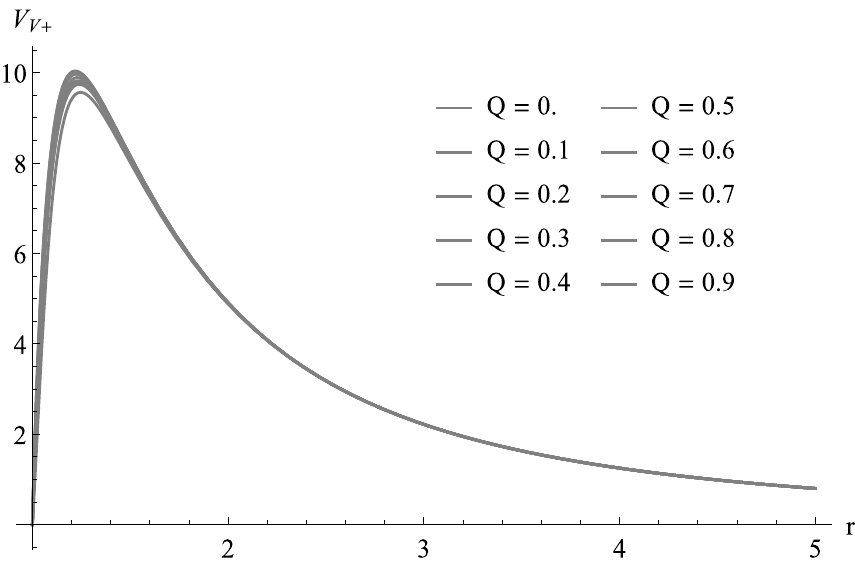}
    \caption{$D=8$}
\end{subfigure}%
\caption{Effective potentials of vector($+$) perturbation in $D=5$ and $D=8$.} \label{VV+58}
\end{figure} 

We illustrated the effective potentials for vector($+$) gravitational perturbations in \autoref{VV+58}. Potentials with a single maximum are plotted in grey. The figure shows that five-dimensional black holes are described by potentials with a single smooth barrier for all $Q$. This leads to a favorable performance of the WKB approach and thus to applicability of the relation between QNMs and GBFs. 

\begin{figure}[h!]
\noindent\begin{subfigure}[b]{0.5\textwidth}
    \centering
    \includegraphics[scale=0.46]{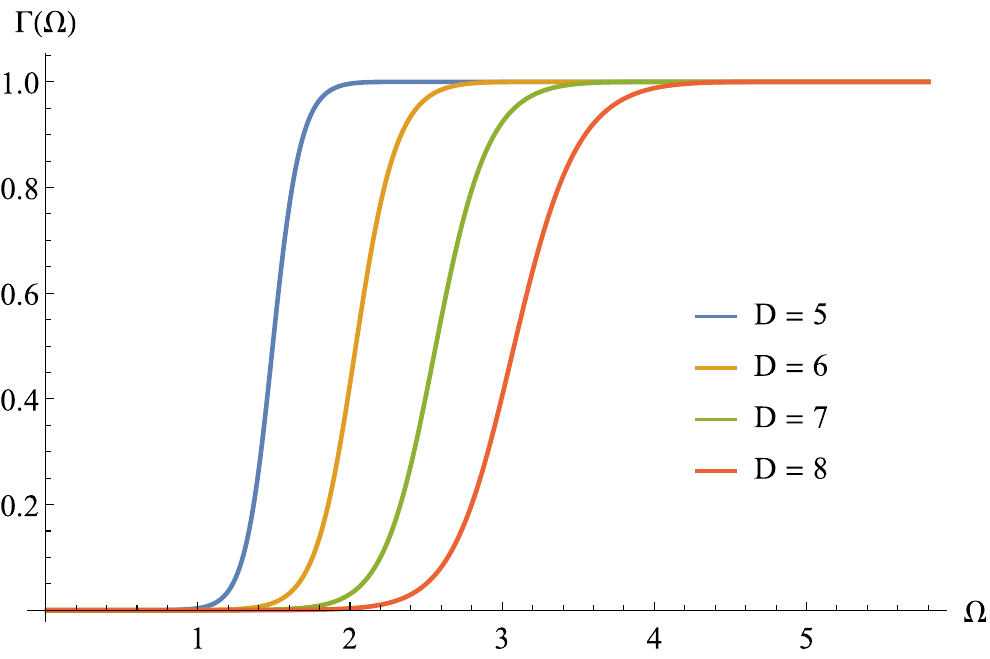}
\end{subfigure}%
\noindent\begin{subfigure}[b]{0.5\textwidth}
    \centering
    \includegraphics[scale=0.45]{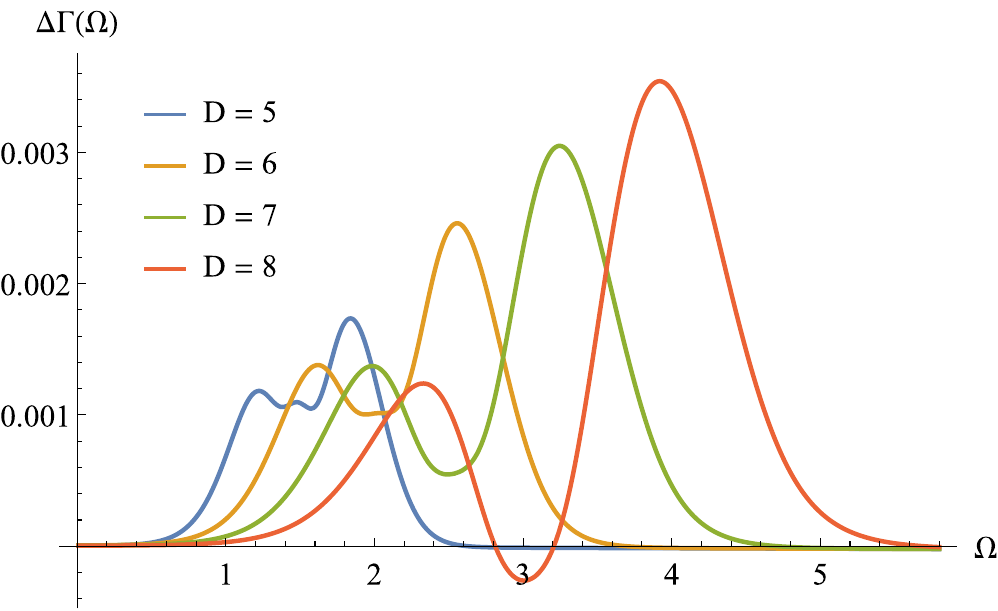}
\end{subfigure}
\caption{GBFs obtained using correspondence for vector($+$) perturbation for $Q=0.6$ (Left) and their differences from the numerically computed GBFs (Right).} \label{gbfV+Q06}
\end{figure}

The single-peak forms in $D=8$ suggest that the correspondence is also valid up to $D=8$. As an example, we calculated the GBFs for $Q=0.6$ in each dimension. Our results for the QNMs and the GBFs obtained by the correspondence are presented in \autoref{qnmV+Q06} and \autoref{gbfV+Q06}, respectively.

\begin{table}[H]
\centering
\begin{tabular}{
|>{\centering\arraybackslash}p{1.5cm}|
 >{\centering\arraybackslash}p{4.5cm}|
 >{\centering\arraybackslash}p{4.5cm}| }

\hline\hline
\multicolumn{3}{|c|}{\textbf{$Q=0.6, \, l=2$}} \\
\hline\hline

$D$ & $n=0$ & $n=1$ \\
\hline
$5$   & $1.483703 - 0.287232 i
$ & $1.418274 - 0.876461 i
$ \\
$6$   & $2.021858 - 0.418583 i
$ & $1.890896 - 1.282078 i
$ \\
$7$   & $2.537245 - 0.538938 i
$ & $2.325498 - 1.652856 i
$ \\
$8$   & $3.040321 - 0.650100 i
$ & $2.734518 - 1.991691 i
$ \\
\hline\hline

\end{tabular}
    \caption{QN frequency for vector($+$) perturbation for $Q=0.6$.} 
\label{qnmV+Q06}
\end{table}

\begin{figure}[h!]
\noindent\begin{subfigure}[b]{0.5\textwidth}
    \centering
    \includegraphics[scale=0.5]{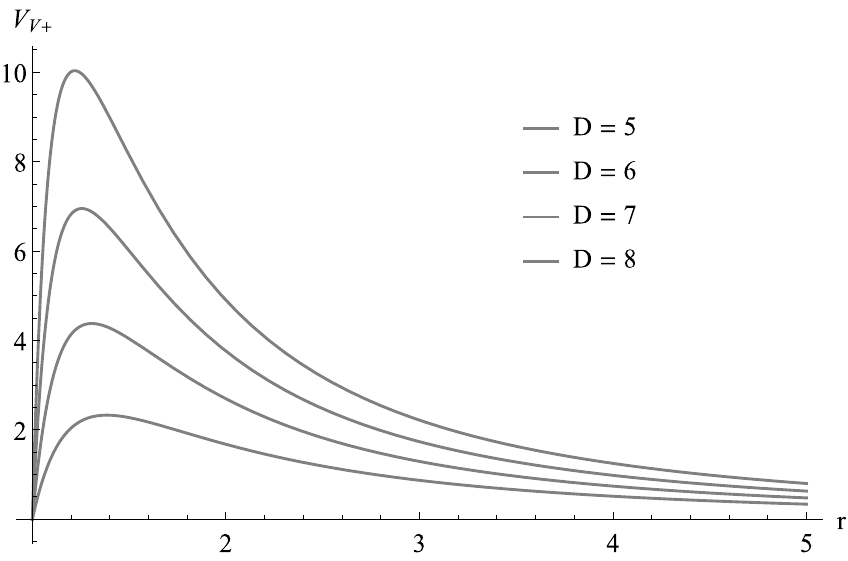}
    \caption{$Q=0.6$}
\end{subfigure}%
\noindent\begin{subfigure}[b]{0.5\textwidth}
    \centering
    \includegraphics[scale=0.5]{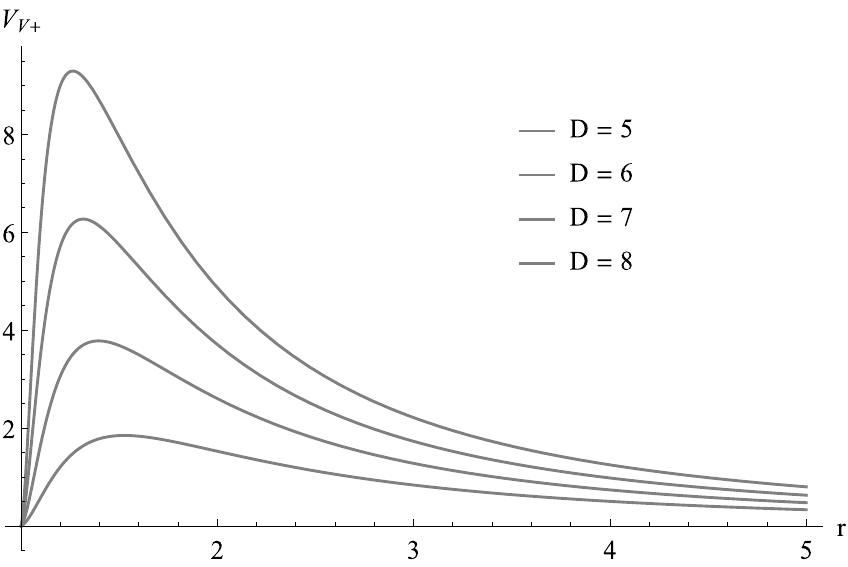}
    \caption{$Q=0.98$}
\end{subfigure}
\caption{Effective potentials of vector($+$) perturbation for $Q=0.6$ and $Q=0.98$.} \label{VV+Q}
\end{figure}

The deviation $\Delta \Gamma(\Omega)$ of the approximate GBFs from the numerical GBFs is very small in $D=5,6,7$, and $8$. Although it increases with the dimension, $\Delta \Gamma(\Omega)$ for $D=8$ remains under $0.004$. Therefore, the correspondence provides a good approximation of GBFs for vector($+$) perturbation in higher dimensions. This result agrees with the form of the potentials for which the WKB approach is applicable, as shown in \autoref{VV+Q}. The accuracy of the correspondence tends to deteriorate with increasing dimension because the WKB approach is less accurate at higher $D$.

\begin{figure}[h!]
\noindent\begin{subfigure}[b]{0.5\textwidth}
    \centering
    \includegraphics[scale=0.57]{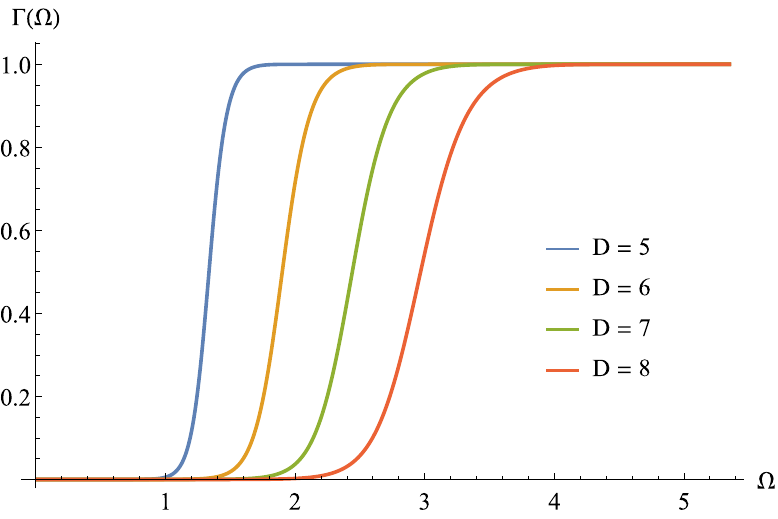}
\end{subfigure}%
\noindent\begin{subfigure}[b]{0.5\textwidth}
    \centering
    \includegraphics[scale=0.57]{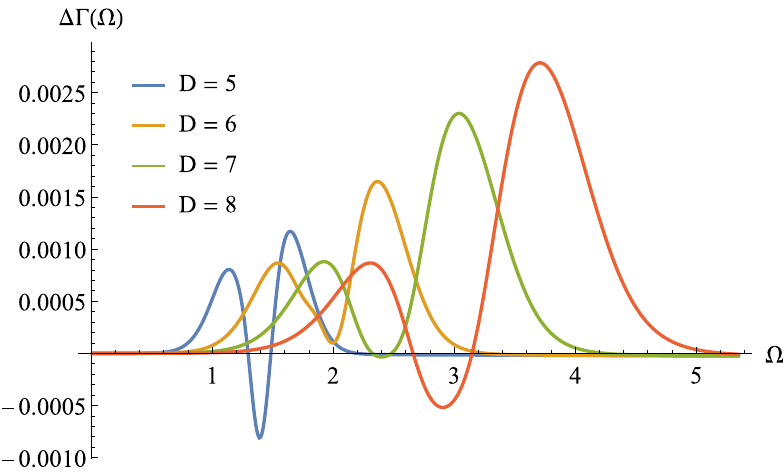}
\end{subfigure}
\caption{GBFs obtained using correspondence for vector($+$) perturbation for $Q=0.98$ (Left) and their differences from the numerically computed GBFs (Right).} \label{gbfV+Q98}
\end{figure}

We also examined the GBFs of near-extreme black holes. At a fixed value $Q=0.98$, the fundamental QNM and the first overtone were obtained by the integration-through-midpoints method, as presented in \autoref{qnmV+Q98}. The GBFs can be determined by substituting the QN frequencies into Eq. \eqref{corr}, and we show their deviations from the numerically computed GBFs in \autoref{gbfV+Q98}. As can be expected from the form of the potentials in \autoref{VV+Q}, the correspondence achieves excellent accuracy in all dimensions considered.

\begin{table}[H]
\centering
\begin{tabular}{
|>{\centering\arraybackslash}p{1.5cm}|
 >{\centering\arraybackslash}p{4.5cm}|
 >{\centering\arraybackslash}p{4.5cm}| }

\hline\hline
\multicolumn{3}{|c|}{\textbf{$Q=0.98, \, l=2$}} \\
\hline\hline

$D$ & $n=0$ & $n=1$ \\
\hline
$5$   & $1.327705 - 0.213719 i
$ & $1.265166 - 0.649368 i$ \\
$6$   & $1.885283 - 0.335215 i
$ & $1.760628 - 1.019818 i
$ \\
$7$   & $2.416801 - 0.448367 i
$ & $2.219944 - 1.362272 i
$ \\
$8$   & $2.933640 - 0.553762 i
$ & $2.657060 - 1.677071 i
$ \\
\hline\hline

\end{tabular}
    \caption{QN frequency for vector($+$) perturbation for $Q=0.98$.} 
\label{qnmV+Q98}
\end{table}

\subsection{Vector($-$) perturbation}
We investigate the correspondence for the vector($-$) perturbation with $V_{V-}$. This type recovers the vector gravitational perturbation of the neutral black hole in the limit $Q \to 0$. We computed the QN frequencies to analytically find the GBFs through the correspondence. The $n=0$ and $n=1$ modes for five-dimensional black holes are listed in \autoref{qnmV-5}. As in scalar($-$) perturbations, the real parts and the magnitudes of imaginary parts of the frequencies decrease monotonically as $Q$ increases.

\begin{table}[H]
\centering
\begin{tabular}{
|>{\centering\arraybackslash}p{1.5cm}|
 >{\centering\arraybackslash}p{4.5cm}|
 >{\centering\arraybackslash}p{4.5cm}| }

\hline\hline
\multicolumn{3}{|c|}{\textbf{$D=5, \, l=2$}} \\
\hline\hline

$Q$ & $n=0$ & $n=1$ \\
\hline
$0$   & $1.134003 - 0.327523 i
$ & $0.947416 - 1.022040 i$ \\
$0.1$   & $1.128587 - 0.325547 i
$ & $0.943677 - 1.016194 i$ \\
$0.2$   & $1.113072 - 0.319749 i
$ & $0.933250 - 0.998977 i$ \\
$0.3$   & $1.089183 - 0.310489 i
$ & $0.917790 - 0.971173 i$ \\
$0.4$   & $1.058771 - 0.298312 i
$ & $0.898394 - 0.933809 i $  \\
$0.5$   & $1.023319 - 0.283981 i
$ & $0.874608 - 0.888414 i$ \\
$0.6$   & $0.983965 - 0.268540 i
$ & $0.844543 - 0.838143 i
$ \\
$0.7$   & $0.941884 - 0.253280 i
$ & $0.807491 - 0.789184 i$ \\
$0.8$   & $0.898664 - 0.239247 i
$ & $0.768716 - 0.746355 i
$ \\
$0.9$   & $0.855832 - 0.226639 i
$ & $0.731365 - 0.707547 i$ \\
\hline\hline

\end{tabular}
    \caption{QN frequency for vector($-$) perturbation in $D=5$.} 
\label{qnmV-5}
\end{table}

\begin{figure}[h!]
\noindent\begin{subfigure}[b]{0.5\textwidth}
    \centering
    \includegraphics[scale=0.47]{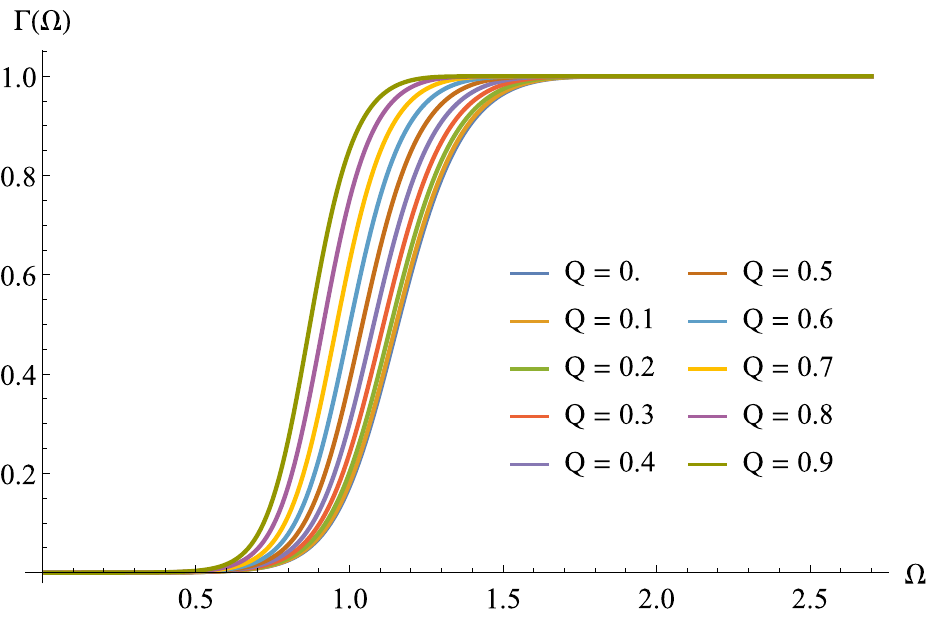}
\end{subfigure}%
\noindent\begin{subfigure}[b]{0.5\textwidth}
    \centering
    \includegraphics[scale=0.42]{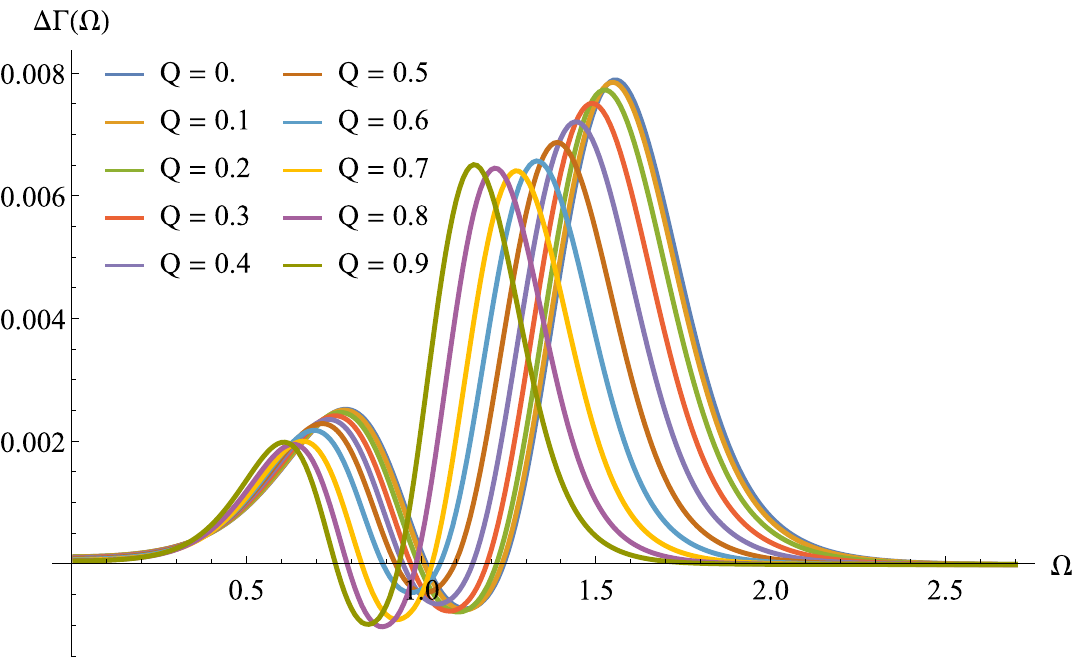}
\end{subfigure}
\caption{GBFs obtained using correspondence for vector($-$) perturbation in $D=5$ (Left) and their differences from the numerically computed GBFs (Right).} \label{gbfV-5}
\end{figure}

\begin{figure}[H] 
\noindent\begin{subfigure}[b]{0.5\textwidth}
    \centering
    \includegraphics[scale=0.51]{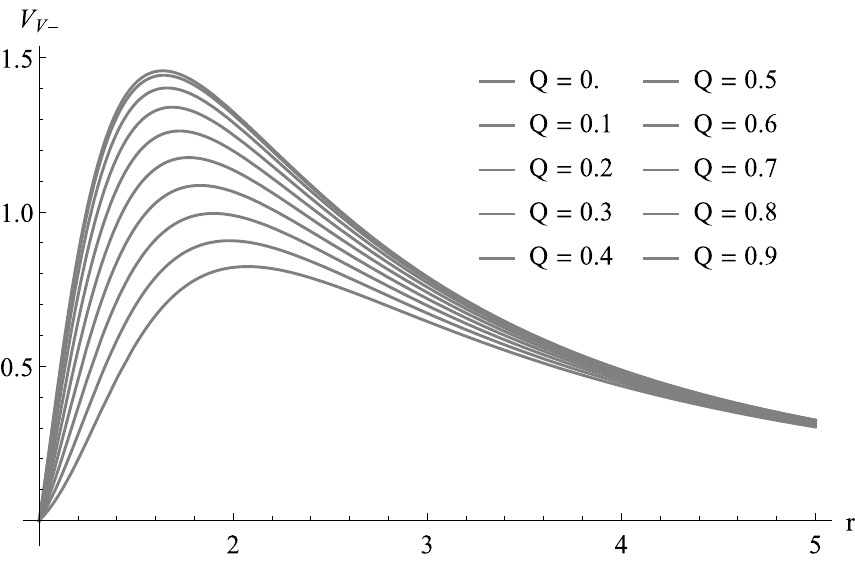}
    \caption{$D=5$}
\end{subfigure}%
\noindent\begin{subfigure}[b]{0.5\textwidth}
    \centering
    \includegraphics[scale=0.51]{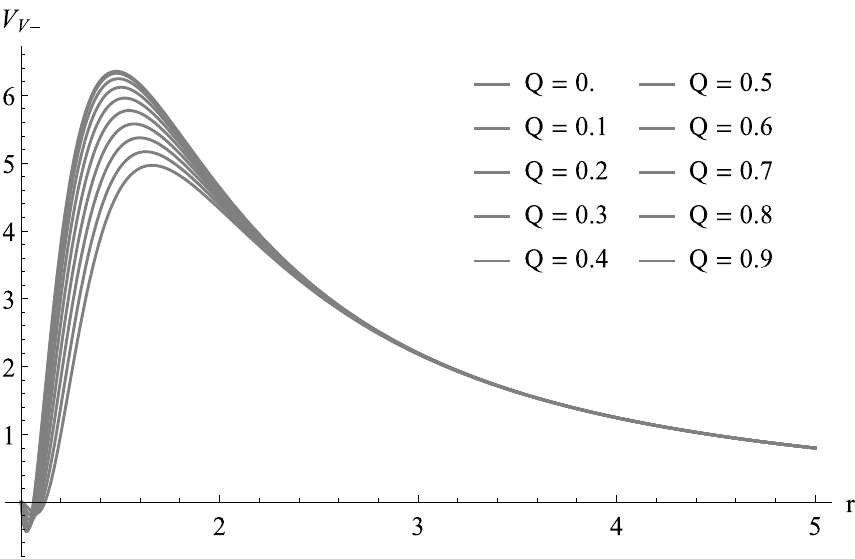}
    \caption{$D=8$}
\end{subfigure}%
\caption{Effective potentials of vector($-$) perturbation in $D=5$ and $D=8$.} \label{VV-D}
\end{figure}

The correspondence \eqref{corr} yields the approximate GBFs plotted in \autoref{gbfV-5}. We also obtained the accurate values using a numerical computation and presented the differences between them and the data obtained by the correspondence. The small differences indicate that the correspondence is valid for all $Q$ in $D=5$. The maximum magnitude of the difference tends to decrease as the charge parameter increases. The reliable performance of the correspondence stems from the effective potential of vector($-$) perturbations, shown in \autoref{VV-D}. The grey graphs show that, for all values of $Q$, the potentials have a single barrier, for which the WKB method performs well. Because the single-barrier form also appears in $D>5$, the correspondence is expected to exhibit good accuracy in higher dimensions.

\begin{figure}[h!]
\noindent\begin{subfigure}[b]{0.5\textwidth}
    \centering
    \includegraphics[scale=0.45]{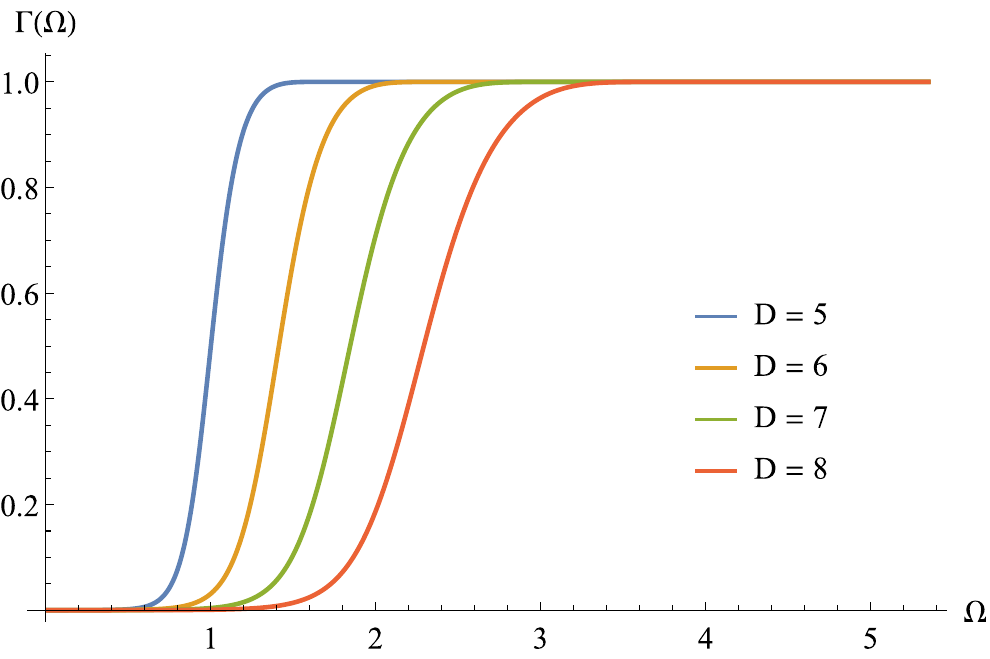}
\end{subfigure}%
\noindent\begin{subfigure}[b]{0.5\textwidth}
    \centering
    \includegraphics[scale=0.45]{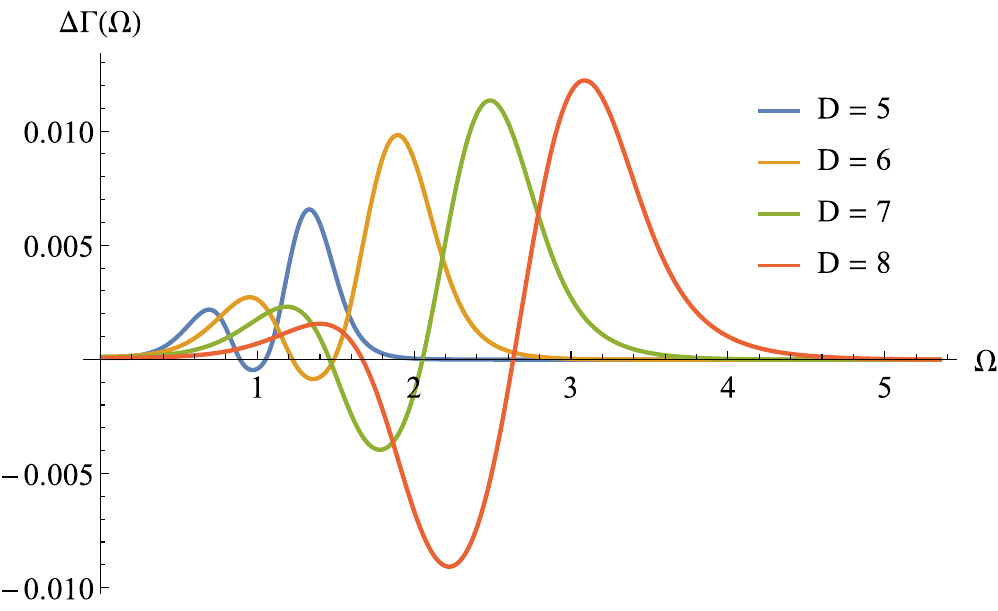}
\end{subfigure}
\caption{GBFs obtained using correspondence for vector($-$) perturbation for $Q=0.6$ (Left) and their differences from the numerically computed GBFs (Right).} \label{gbfV-Q06}
\end{figure}

We present the analysis for a fixed $Q$ as a representative example. The QNMs are computed for black holes with $Q=0.6$ and given in \autoref{qnmV-Q06}. Upon substituting them into Eq. \eqref{corr}, we obtain the GBFs. \autoref{gbfV-Q06} shows the GBFs found from the correspondence and their deviation from the GBFs obtained numerically. As expected, the correspondence behaves well for vector($-$) perturbations in higher dimensions. We also observe that its precision tends to decrease with increasing $D$.

\begin{table}[H]
\centering
\begin{tabular}{
|>{\centering\arraybackslash}p{1.5cm}|
 >{\centering\arraybackslash}p{4.5cm}|
 >{\centering\arraybackslash}p{4.5cm}| }

\hline\hline
\multicolumn{3}{|c|}{\textbf{$Q=0.6, \, l=2$}} \\
\hline\hline

$D$ & $n=0$ & $n=1$ \\
\hline
$5$   & $0.983965 - 0.268540 i
$ & $0.844543 - 0.838143 i
$ \\
$6$   & $1.380946 - 0.408556 i
$ & $1.116060 - 1.293708 i
$ \\
$7$   & $1.794773 - 0.539415 i
$ & $1.382503 - 1.699780 i
$ \\
$8$   & $2.219703 - 0.659717 i
$ & $1.630004 - 2.046938 i
$ \\
\hline\hline

\end{tabular}
    \caption{QN frequency for vector($-$) perturbation for $Q=0.6$.} 
\label{qnmV-Q06}
\end{table}

\begin{figure}[h!]
\noindent\begin{subfigure}[b]{0.5\textwidth}
    \centering
    \includegraphics[scale=0.52]{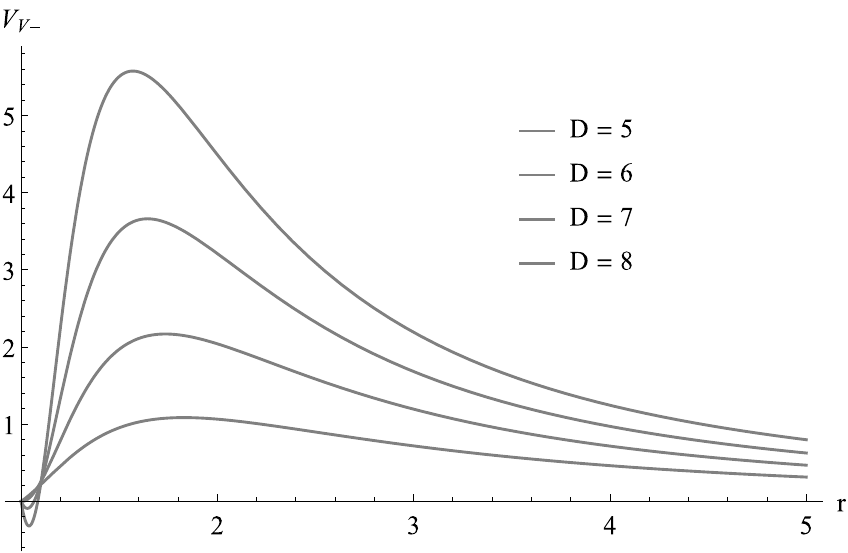}
    \caption{$Q=0.6$}
\end{subfigure}%
\noindent\begin{subfigure}[b]{0.5\textwidth}
    \centering
    \includegraphics[scale=0.52]{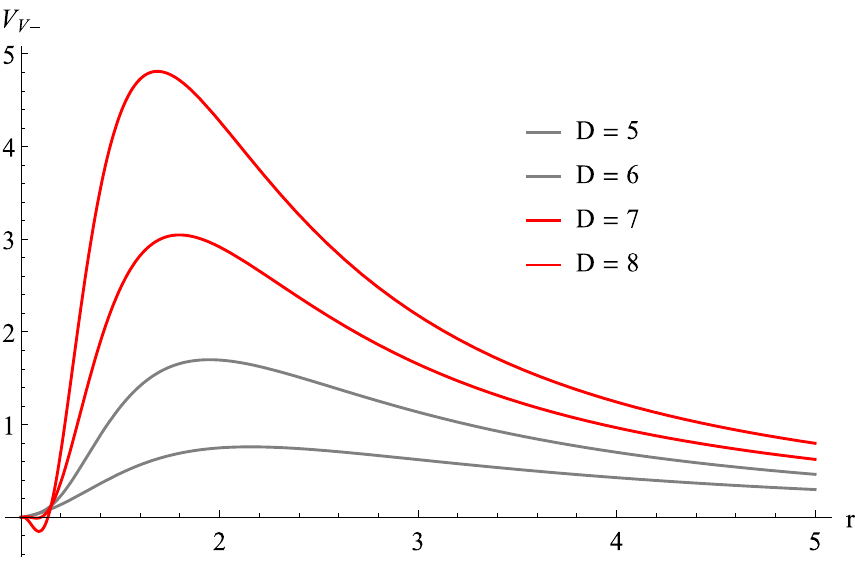}
    \caption{$Q=0.98$}
\end{subfigure}
\caption{Effective potentials of vector($-$) perturbation for $Q=0.6$ and $Q=0.98$.} \label{VV-Q}
\end{figure}

We plotted $V_{V-}$ at $Q=0.6$ and $Q=0.98$ for each dimension in \autoref{VV-Q}. The graphs are shown in grey if the potential has a single peak and in red otherwise. For $Q=0.6$, all cases have a single peak, consistent with the result that the correspondence provides the GBFs with high accuracy. However, double peaks appear for $D>6$ in the near-extreme regime. We tested the accuracy of the correspondence for near-extreme black holes with $Q=0.98$. The QN frequencies listed in \autoref{qnmV-Q098} were obtained by introducing the appropriate midpoints in series \eqref{news} and increasing the depth of the continued fraction.

\begin{table}[h!]
\centering
\begin{tabular}{
|>{\centering\arraybackslash}p{1.5cm}|
 >{\centering\arraybackslash}p{4.5cm}|
 >{\centering\arraybackslash}p{4.5cm}| }

\hline\hline
\multicolumn{3}{|c|}{\textbf{$Q=0.98, \, l=2$}} \\
\hline\hline

$D$ & $n=0$ & $n=1$ \\
\hline
$5$   & $0.822458 - 0.217472 i
$ & $0.702638 - 0.679079 i$ \\
$6$   & $1.220852 - 0.354466 i
$ & $0.979881 - 1.119980 i
$ \\
$7$   & $1.633724 - 0.484140 i
$ & $1.245666  - 1.523965 i
$ \\
$8$   & $2.056894 - 0.604571 i
$ & $1.493496 - 1.873464 i
$ \\
\hline\hline

\end{tabular}
    \caption{QN frequency for vector($-$) perturbation for $Q=0.98$.} 
\label{qnmV-Q098}
\end{table}

\begin{figure}[h!]
\noindent\begin{subfigure}[b]{0.5\textwidth}
    \centering
    \includegraphics[scale=0.5]{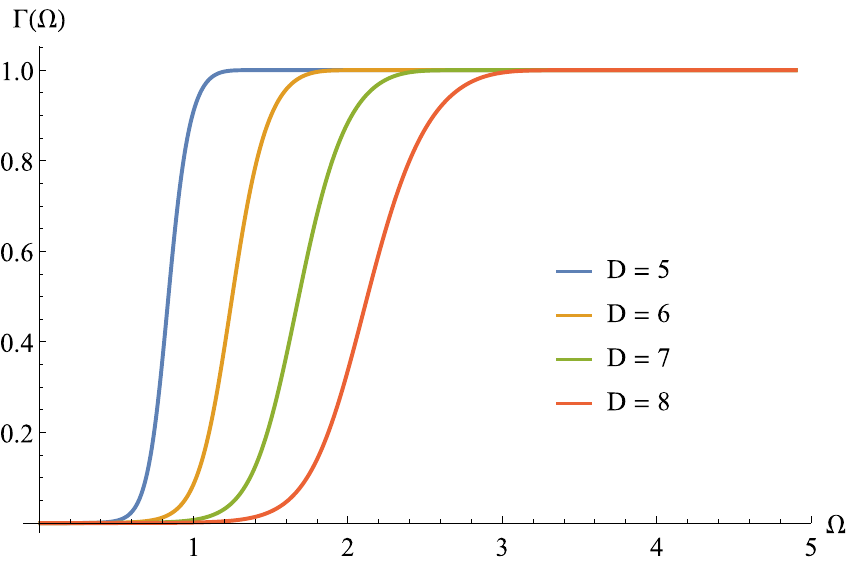}
\end{subfigure}%
\noindent\begin{subfigure}[b]{0.5\textwidth}
    \centering
    \includegraphics[scale=0.52]{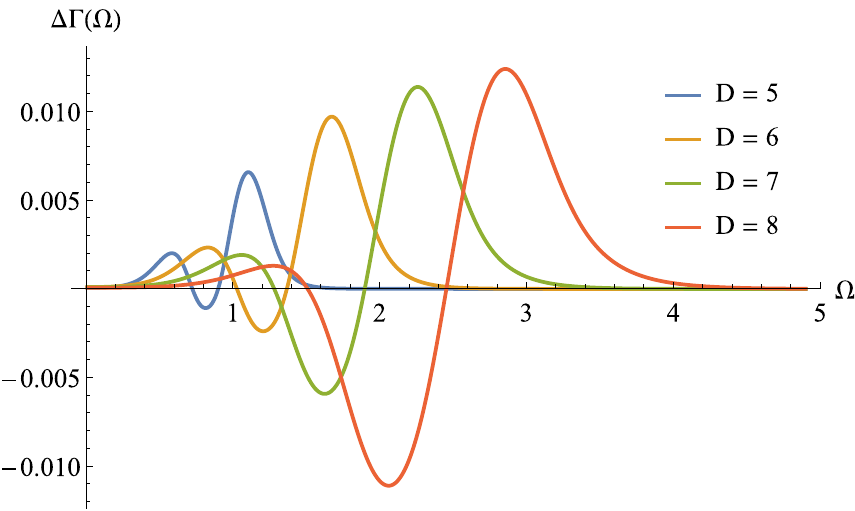}
\end{subfigure}
\caption{GBFs obtained using correspondence for vector($-$) perturbation for $Q=0.98$ (Left) and their differences from the numerically computed GBFs (Right).} \label{gbfV+Q098}
\end{figure}

\autoref{gbfV+Q098} illustrates the GBFs found by the correspondence using the QNMs and the differences of these GBFs from the values obtained from the numerical method. Although the potential develops a double-peak structure in $D=7$ and $D=8$, the correspondence achieves good accuracy and exhibits continuous variation with dimension. This is presumably because the second peak occurring near the event horizon is negligibly small, making the potential effectively behave as a single barrier. These results demonstrate that the correspondence for the vector($-$) perturbation performs well over the range of charge considered in higher dimensions.

\section{Correspondence in tensor-type perturbations}
Finally, the relationship between QNMs and GBFs is examined for tensor perturbations, described by the effective potential $V_T$ \eqref{VT}. For non-extreme black holes with charge parameter $0\le Q <1$, the QNMs and the GBFs are obtained for multipole $l=2$. We start by computing the fundamental and first overtone modes $\omega_0$ and $\omega_1$ for $D=5$, as given in \autoref{qnmT5}.

\begin{table}[H]
\centering
\begin{tabular}{
|>{\centering\arraybackslash}p{1.5cm}|
 >{\centering\arraybackslash}p{4.5cm}|
 >{\centering\arraybackslash}p{4.5cm}| }

\hline\hline
\multicolumn{3}{|c|}{\textbf{$D=5, \, l=2$}} \\
\hline\hline

$Q$ & $n=0$ & $n=1$ \\
\hline
$0$   & $1.510567 - 0.357537 i
$ & $1.392721 - 1.104557 i$ \\
$0.1$   & $1.506872 - 0.355275 i
$ & $1.390854 - 1.097336 i$ \\
$0.2$   & $1.495752 - 0.348578 i
$ & $1.384951 - 1.075946 i$ \\
$0.3$   & $1.477114 - 0.337732 i
$ & $1.374100 - 1.041246 i$ \\
$0.4$   & $1.450835 - 0.323260 i
$ & $1.356791 - 0.994819 i $  \\
$0.5$   & $1.416838 - 0.305998 i
$ & $1.331019 - 0.939278 i$ \\
$0.6$   & $1.375258 - 0.287172 i
$ & $1.294781 - 0.878842 i
$ \\
$0.7$   & $1.326764 - 0.268377 i
$ & $1.247754 - 0.819702 i$ \\
$0.8$   & $1.272902 - 0.251178 i
$ & $1.193731 - 0.767542 i
$ \\
$0.9$   & $1.215894 - 0.236429 i
$ & $1.137804 - 0.723438 i$ \\
\hline\hline

\end{tabular}
    \caption{QN frequency for tensor perturbation in $D=5$.} 
\label{qnmT5}
\end{table}

\begin{figure}[h!]
\noindent\begin{subfigure}[b]{0.5\textwidth}
    \centering
    \includegraphics[scale=0.5]{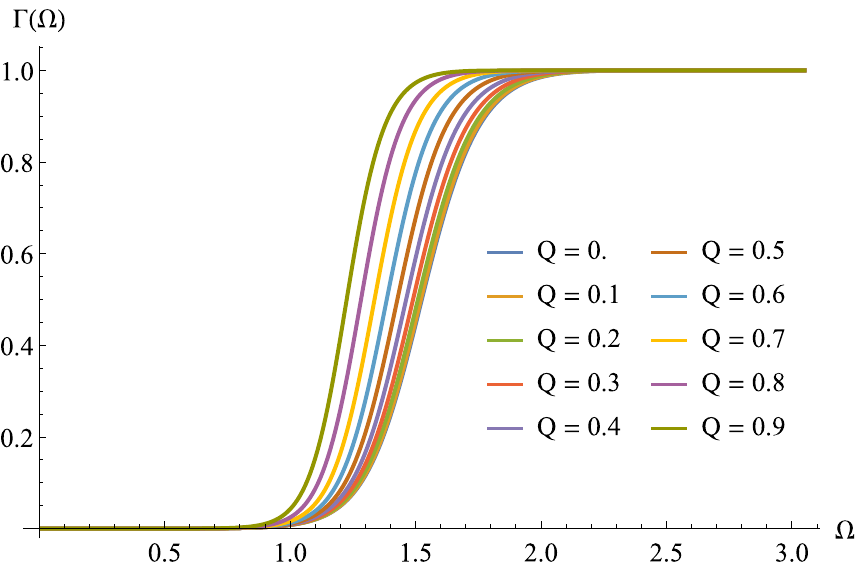}
\end{subfigure}%
\noindent\begin{subfigure}[b]{0.5\textwidth}
    \centering
    \includegraphics[scale=0.45]{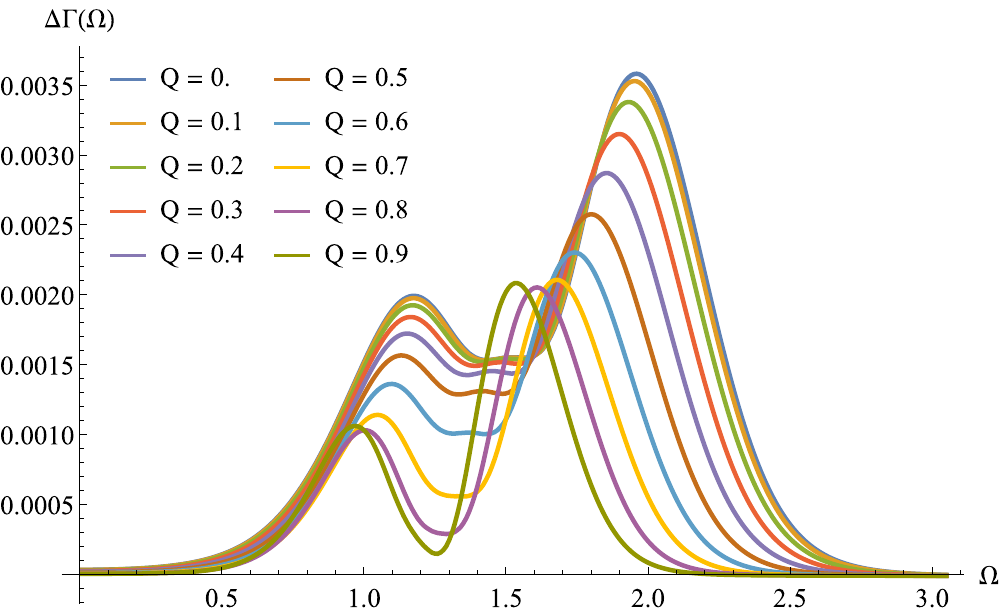}
\end{subfigure}
\caption{GBFs obtained using correspondence for tensor perturbation in $D=5$ (Left) and their differences from the numerically computed GBFs (Right).} \label{gbfT5}
\end{figure}

As observed in the cases of scalar($-$) and vector($-$) perturbations, the real parts and the magnitudes of the imaginary parts decrease monotonically with growing $Q$. We checked that $\omega_0$ is consistent with the values presented in \cite{Konoplya:2007jv}. The QN frequencies are used to determine the GBFs through the correspondence \eqref{corr}. \autoref{gbfT5} illustrates the resulting GBFs and their discrepancies from the accurate GBFs computed numerically. We observe that the discrepancies remain small, implying that the correspondence holds for tensor perturbations in $D=5$. Because the maximum of $|\Delta \Gamma(\Omega)|$ decreases as $Q$ rises, the correspondence tends to be more accurate for a higher charge.

\begin{figure}[h!] 
\noindent\begin{subfigure}[b]{0.5\textwidth}
    \centering
    \includegraphics[scale=0.5]{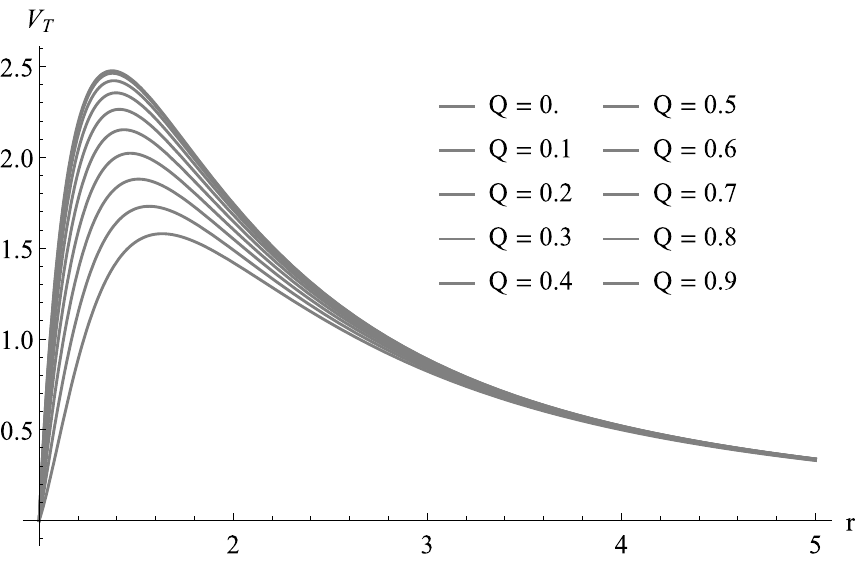}
    \caption{$D=5$}
\end{subfigure}%
\noindent\begin{subfigure}[b]{0.5\textwidth}
    \centering
    \includegraphics[scale=0.5]{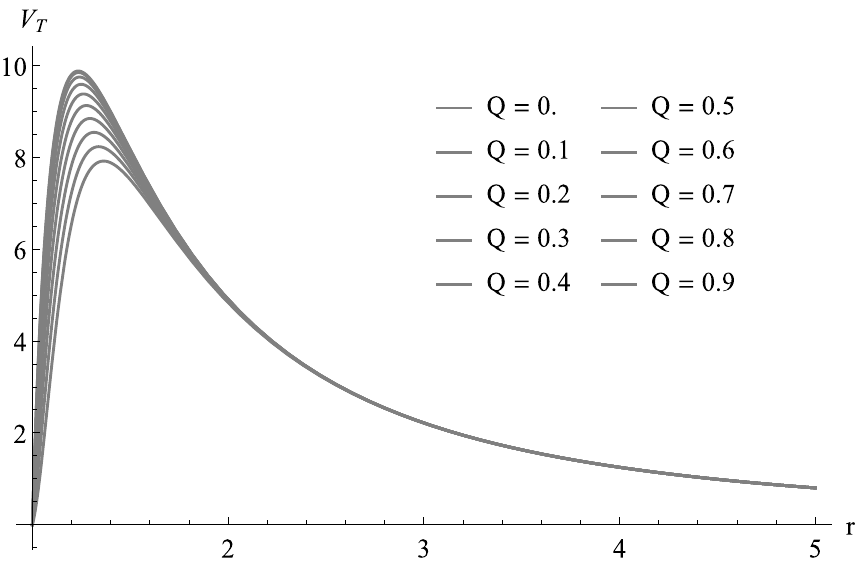}
    \caption{$D=8$}
\end{subfigure}%
\caption{Effective potentials of tensor perturbation in $D=5$ and $D=8$.} \label{VTD}
\end{figure}

\begin{figure}[h!]
\noindent\begin{subfigure}[b]{0.5\textwidth}
    \centering
    \includegraphics[scale=0.45]{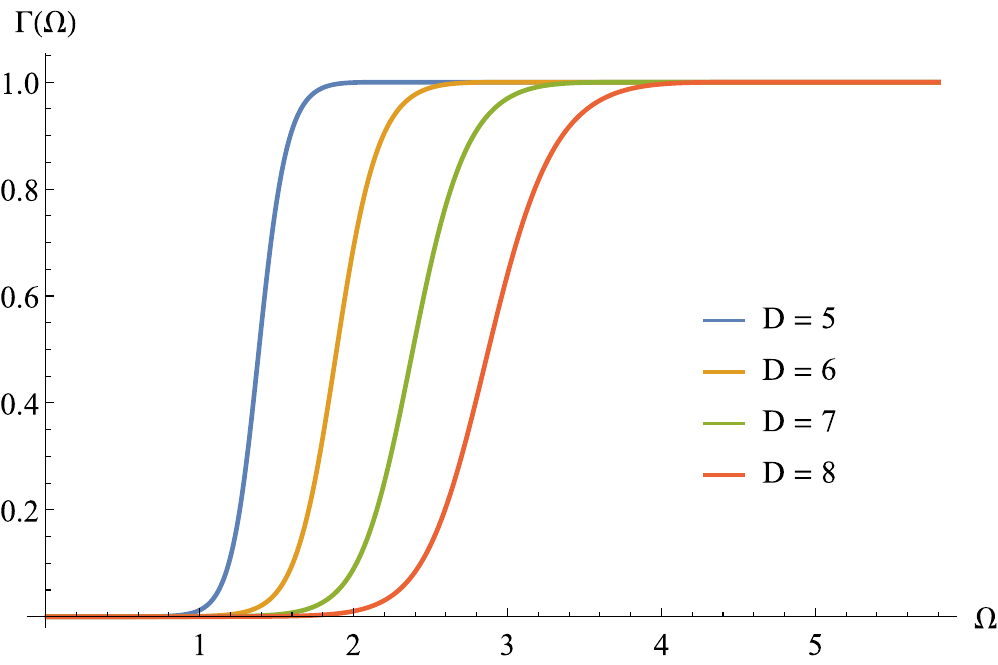}
\end{subfigure}%
\noindent\begin{subfigure}[b]{0.5\textwidth}
    \centering
    \includegraphics[scale=0.45]{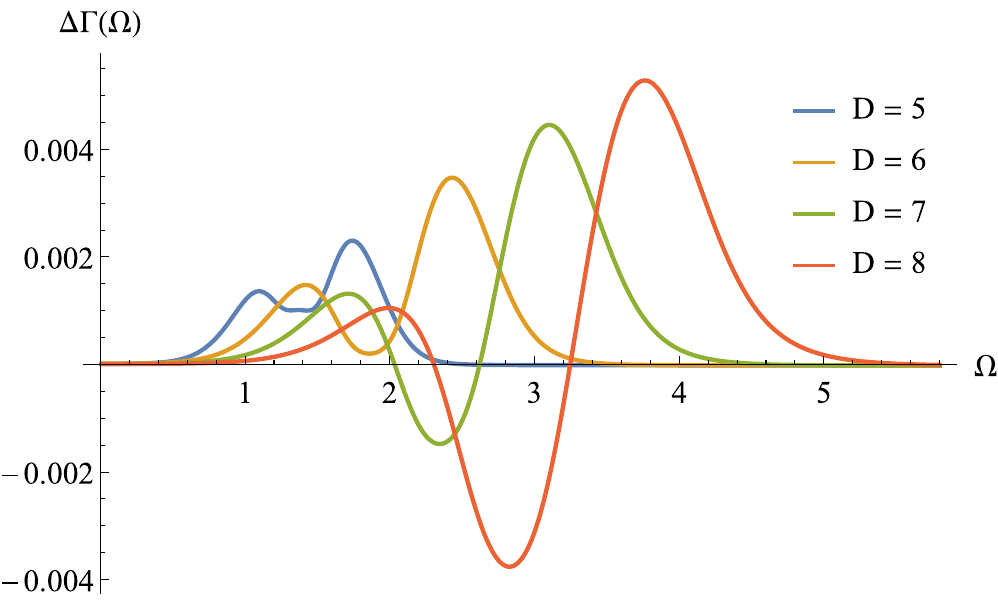}
\end{subfigure}
\caption{GBFs obtained using correspondence for tensor perturbation for $Q=0.6$ (Left) and their differences from the numerically computed GBFs (Right).} \label{gbfTQ06}
\end{figure}

In the plots of the effective potential $V_T$ in \autoref{VTD}, all cases exhibit a single barrier. This form leads to good performance of the WKB approximation employed to construct the correspondence. The potentials in $D=8$ also have a smooth positive barrier for all $Q$. To explicitly test the correspondence in higher dimensions, we computed the QNMs and GBFs for $Q=0.6$ in each $D$. The results are presented in \autoref{qnmTQ06} and \autoref{gbfTQ06}, respectively.

\begin{table}[h!]
\centering
\begin{tabular}{
|>{\centering\arraybackslash}p{1.5cm}|
 >{\centering\arraybackslash}p{4.5cm}|
 >{\centering\arraybackslash}p{4.5cm}| }

\hline\hline
\multicolumn{3}{|c|}{\textbf{$Q=0.6, \, l=2$}} \\
\hline\hline

$D$ & $n=0$ & $n=1$ \\
\hline
$5$   & $1.375258 - 0.287172 i
$ & $1.294781 - 0.878842 i$ \\
$6$   & $1.869398 - 0.422779 i
$ & $1.698688 - 1.299404 i$ \\
$7$   & $2.350024 - 0.548305 i
$ & $2.063916 - 1.684319 i$  \\
$8$   & $2.824234-0.664605 i$ & $2.401771-2.029485 i$ \\

\hline\hline

\end{tabular}
    \caption{QN frequency for tensor perturbation for $Q=0.6$.} 
\label{qnmTQ06}
\end{table}

\begin{figure}[h!]
\noindent\begin{subfigure}[b]{0.5\textwidth}
    \centering
    \includegraphics[scale=0.52]{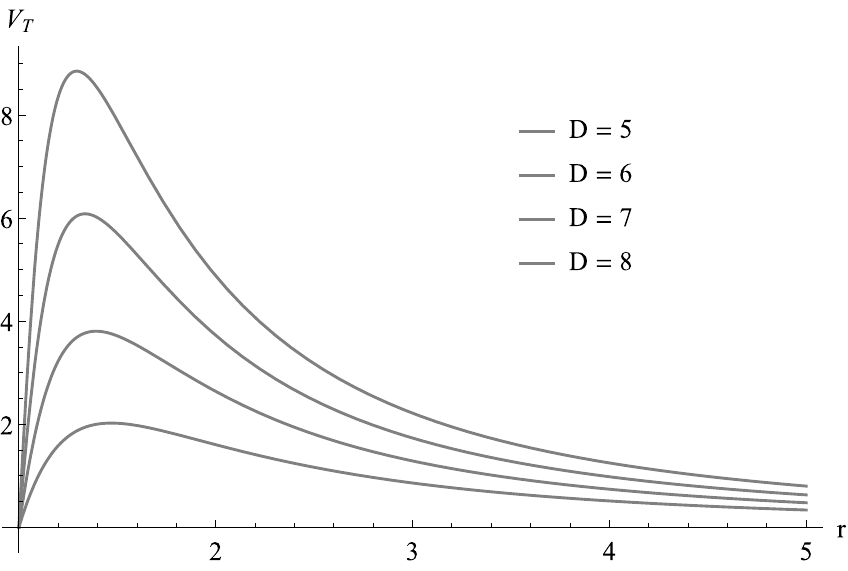}
    \caption{$Q=0.6$}
\end{subfigure}%
\noindent\begin{subfigure}[b]{0.5\textwidth}
    \centering
    \includegraphics[scale=0.52]{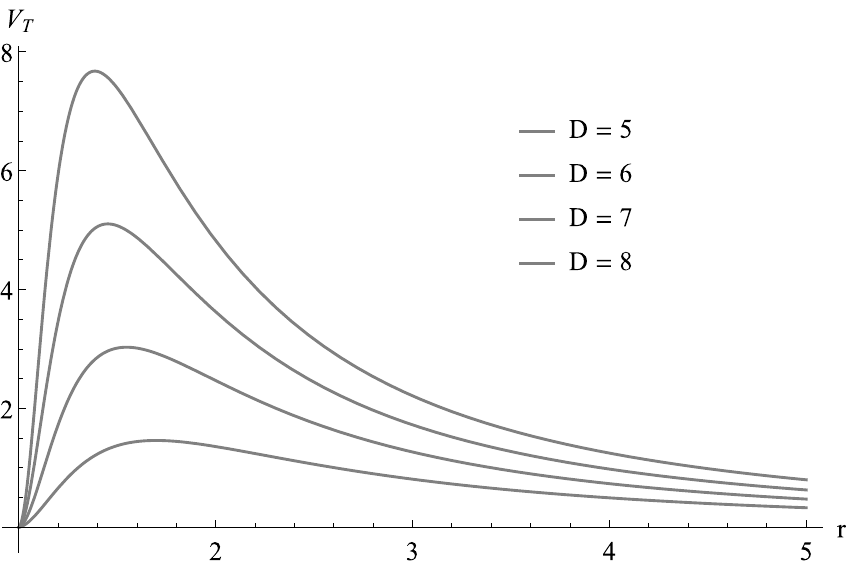}
    \caption{$Q=0.98$}
\end{subfigure}
\caption{Effective potentials of tensor perturbation for $Q=0.6$ and $Q=0.98$.} \label{VTQ}
\end{figure}

The discrepancies between the GBFs determined by the correspondence and those obtained by the numerical method are small for $D=5,6,7,$ and $8$, suggesting that the correspondence is valid with high precision. This is attributed to the single-peak form of the potentials, as presented in \autoref{VTQ} for $Q=0.6$. Although the accuracy decreases as the dimension increases, it remains at a reliable level in all dimensions considered.

\begin{figure}[h!]
\noindent\begin{subfigure}[b]{0.5\textwidth}
    \centering
    \includegraphics[scale=0.55]{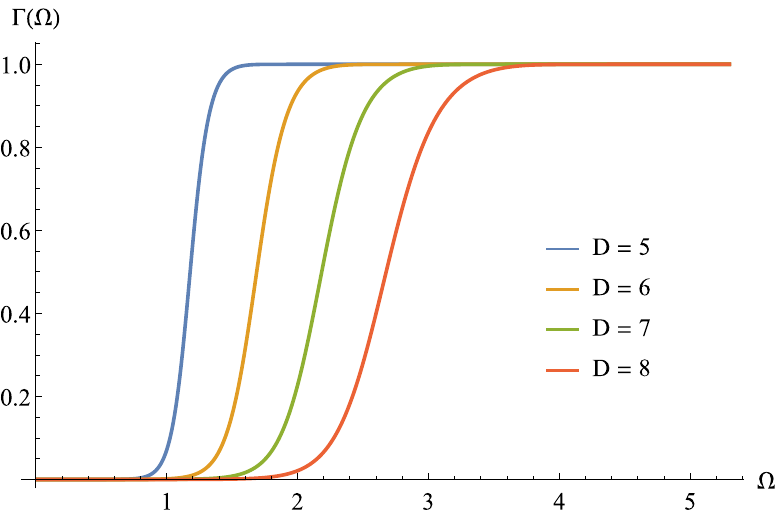}
\end{subfigure}%
\noindent\begin{subfigure}[b]{0.5\textwidth}
    \centering
    \includegraphics[scale=0.57]{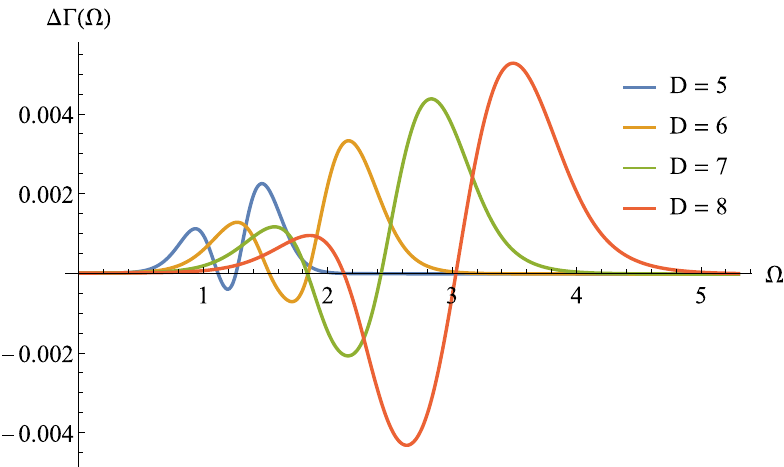}
\end{subfigure}
\caption{GBFs obtained using correspondence for tensor perturbation for $Q=0.98$ (Left) and their differences from the numerically computed GBFs (Right).} \label{gbfTQ098}
\end{figure}

Furthermore, we calculated the GBFs of near-extreme black holes. As shown in the right panel of \autoref{VTQ}, the potential $V_T$ features a single positive peak for the near-extreme cases. We numerically computed the fundamental mode and the first overtone using the integration-through-midpoints method. The frequencies for each dimension are presented in \autoref{qnmTQ098}. By substituting them into the correspondence \eqref{corr}, we obtained the GBFs for a given real frequency $\Omega$. \autoref{gbfTQ098} exhibits the deviations of the GBFs determined by the correspondence from the numerically obtained GBFs. For tensor perturbations of near-extreme black holes, the deviations remain low in higher dimensions, indicating that the correspondence enables the calculation of GBFs with remarkable precision.

\begin{table}[h!]
\centering
\begin{tabular}{
|>{\centering\arraybackslash}p{1.5cm}|
 >{\centering\arraybackslash}p{4.5cm}|
 >{\centering\arraybackslash}p{4.5cm}| }

\hline\hline
\multicolumn{3}{|c|}{\textbf{$Q=0.98, \, l=2$}} \\
\hline\hline

$D$ & $n=0$ & $n=1$ \\
\hline
$5$   & $1.169484 - 0.226412 i
$ & $1.093613 - 0.693164 i$ \\
$6$   & $1.667444 - 0.360828 i
$ & $1.505671 - 1.109493 i
$ \\
$7$   & $2.150961 - 0.487241 i
$ & $1.879301 - 1.496703 i
$ \\
$8$   & $2.627234 - 0.605116 i
$ & $2.225166 - 1.846598 i
$ \\
\hline\hline

\end{tabular}
    \caption{QN frequency for tensor perturbation for $Q=0.98$.} 
\label{qnmTQ098}
\end{table}

We found that the correspondence performs well for tensor perturbations in all dimensions and for all black-hole parameters considered. Because the wave equation for perturbations of a massless scalar field in this background is identical to that for tensor gravitational perturbations, the correspondence should also hold with high precision for the scalar field perturbations.

\section{Conclusions}
We studied the correspondence between QNMs and GBFs for gravitational perturbations of charged black holes in higher dimensions. The correspondence, formulated using the WKB approach, provides an approximation for the GBFs beyond the eikonal regime. By considering scalar, vector, and tensor types of perturbation of the higher-dimensional Reissner–Nordstr\"{o}m black hole, we examined the validity of the correspondence for a low multipole $l=2$ across a range of spacetime dimensions and black-hole charges.

To determine the GBFs through the correspondence, we calculated accurate QNMs by employing the continued fraction method. In some cases of scalar gravitational perturbations, the singularity structure in the Frobenius series required the integration-through-midpoints method. We also employed it for near-extreme black holes to improve the convergence of the series. The correspondence formula enables us to analytically obtain the approximate GBFs from the QN frequencies of the fundamental mode and the first overtone. We calculated the deviations of these approximate values from the numerically obtained GBFs to evaluate the precision of the correspondence. 

The scalar and vector perturbations are divided into ($+$) and ($-$) types due to the coupling of gravitational and electromagnetic perturbations in charged black holes. The correspondence for scalar($+$) perturbations yielded the GBFs with high accuracy for all dimensions and charges considered. This result agrees with the good performance of the WKB approach for the single-peak effective potential of the scalar($+$) type. The accuracy of the correspondence improves for higher charges and lower dimensions. The latter is attributed to the higher precision of the WKB approach at lower dimensions. 

Scalar($-$) perturbations exhibited behavior distinct from that of the other types of perturbation. The GBFs obtained by the correspondence achieved a reasonable accuracy for all $Q$ considered in $D=5$ and $D=6$, where the potentials have a single peak. In $D=7$, the double-peak potential begins to appear, and the accuracy of the correspondence deteriorated for low and intermediate values of $Q$. However, the accuracy increased for high $Q$, contrary to its expected degradation. This suggests that the correspondence can perform reliably even when the potential falls outside the assumption of the standard WKB method. Furthermore, the GBFs for a fixed charge in $D=8$ are more accurate than those in $D=7$. This behavior occurred only in the scalar($-$) type.

For vector($+$) and vector($-$) perturbations, the GBFs obtained using the correspondence attained sufficient accuracy for the entire set of charges and dimensions. As in the scalar($+$) type, the accuracy improved with increasing charge. For a fixed charge, it decreased as the dimension increased. Although the potential for vector($-$) perturbations of near-extreme black holes has two peaks in $D \ge 7$, the correspondence remains valid. 

Finally, the correspondence for tensor perturbations showed remarkable accuracy under all investigated conditions. The potentials for the tensor type possess a single positive barrier for all $Q$ in higher dimensions, including near-extreme values. This form is favorable for the WKB approach and thus for the correspondence between QNMs and GBFs. As in the cases of the scalar($+$), vector($+$), and vector($-$) perturbations, the accuracy of the correspondence improves for higher charges and lower dimensions. The results for tensor gravitational perturbations are also applicable to massless scalar field perturbations, as both are governed by an identical equation.  

Although the performance of the correspondence between QNMs and GBFs is expected to be closely related to that of the WKB method, our results indicate that the correspondence behaves well even when the assumption underlying the method is not strictly satisfied. This finding suggests a broader applicability of the correspondence than indicated by the conditions of its WKB-based derivation.

\vspace{10pt} 

\noindent{\bf Acknowledgments}

\noindent This research was supported by Basic Science Research Program through the National Research Foundation of Korea (NRF) funded by the Ministry of Education (NRF-2022R1I1A2063176) and the Dongguk University Research Fund of 2026.\\

\bibliographystyle{bibstyle}
\bibliography{ref}

\end{document}